\documentclass[lettersize,journal]{IEEEtran}

\usepackage{amsmath,amsfonts}
\usepackage{algorithmic}
\usepackage{algorithm}
\usepackage{array}

\usepackage{graphicx}
\usepackage[caption=false,font=footnotesize]{subfig}

\usepackage{textcomp}
\usepackage{stfloats}
\usepackage{url}
\usepackage{verbatim}
\usepackage{cite}

\usepackage{booktabs}
\usepackage{colortbl}
\usepackage[hidelinks]{hyperref}

\begin{document}

\title{From Objectives to Applications: Aligning Architectural Biases in Audio Self-Supervised Learning}

\author{Kele Xu, Yulu Fang, Boda Zhou, Yulin Sun, Qisheng Xu, Qiya Song, Jin Zhang, Cheng Yang, Huaimin Wang
        % <-this % stops a space
\thanks{Corresponding authors: Yunlin Sun (sunyulin@nudt.edu.cn) and Qisheng Xu (qishengxu@nudt.edu.cn).}}

\maketitle

\begin{abstract}
% This paper presents a systematic framework for understanding self-supervised learning in audio signal processing, tracing its trajectory from early heuristic pretext tasks to the emergence of Native Large Audio Language Models and omni-modal foundation systems. We establish a multidimensional taxonomy that organizes methodologies into five core paradigms: auxiliary tasks, contrastive discrimination, generative reconstruction, discrete semantic prediction, and multimodal alignment. Central to our analysis is the explicit alignment between pre-training objectives and architectural inductive biases, encompassing the locality of convolutional neural networks, the global contextual modeling of Transformers, and the linear-time efficiency of modern State Space Models such as Mamba. Furthermore, we synthesize the representational impact of these methods across diverse domains, including speech processing, environmental sound analysis, and healthcare bioacoustics. Special emphasis is placed on recent breakthroughs, including Neural Codec Language Models and Flow Matching paradigms. The work also provides a synthesis of standardized benchmark suites and evaluation metrics. Finally, we identify the ``perceptual bottleneck'' and security vulnerabilities as critical challenges, while outlining future directions toward physics-informed audio-visual intelligence.
This paper examines audio self-supervised learning (SSL) through the alignment between pretraining objectives, architectural inductive biases, and downstream applications. Rather than treating SSL methods as a chronological sequence of pretext tasks or model families, we ask how different supervisory signals shape the representations that models are expected to learn. The discussion is organized around five paradigms: auxiliary tasks, contrastive learning, generative reconstruction, discrete token prediction, and multimodal alignment. These objectives place different demands on the model, from local structural sensitivity and contrastive invariance to contextual inference, discrete semantic abstraction, and multimodal grounding. We relate these demands to the biases of CNNs, recurrent and State Space Models, Transformers, and hybrid architectures, showing how local acoustic compression, sequential state propagation, content-dependent global routing, and local--global integration support different forms of audio SSL. The same view is then used to interpret downstream applications in speech processing, environmental sound analysis, music information retrieval, medical and bioacoustic analysis, and multimodal audio understanding as practical tests of whether learned representations and architectural choices generalize across domains. We also review benchmark protocols and open challenges, including tokenization bottlenecks, long-context efficiency, robustness, and secure multimodal deployment, and discuss how codec-based tokenization and audio-language modeling extend this objective--architecture--application pipeline.
The accompanying repository is released at \url{https://github.com/colaudiolab/Awesome-Self-Supervised-Audio-Learning}.
\end{abstract}


\section{Introduction}
\label{sec:introduction}
\begin{figure}[!t]
    \centering
    \includegraphics[width=\linewidth]{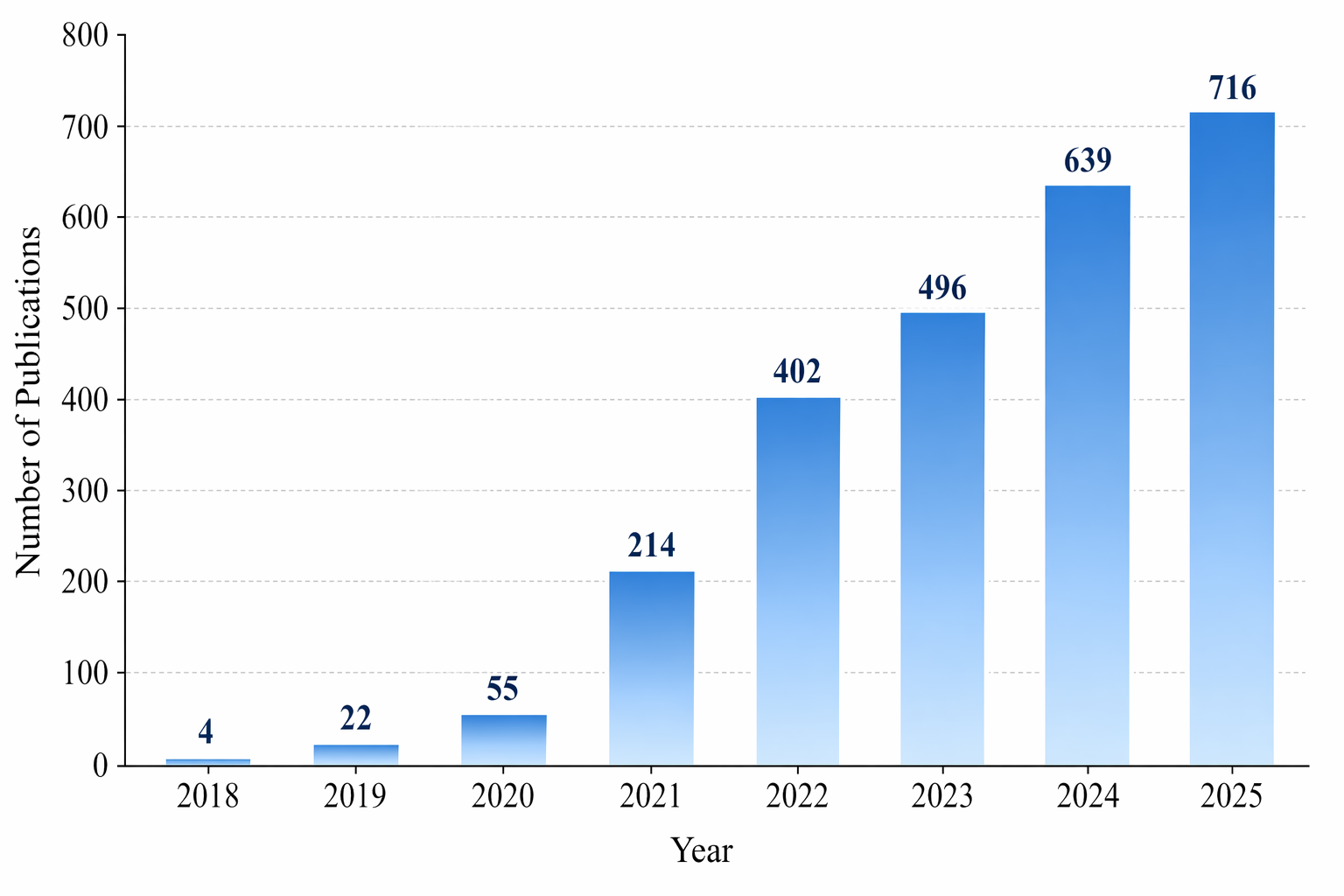}
    \caption{Evolution of research activity in self-supervised audio representation learning. Annual publication counts indicate the growing scholarly attention to audio SSL, with a marked increase in recent years.}
    \label{fig:search}
\end{figure}

\IEEEPARstart{D}{eep} supervised learning has substantially advanced audio signal processing~\cite{LeCun2015DeepLearning,purwins2019deep,richard2023audio}. However, its success is largely predicated on the availability of large-scale labeled datasets~\cite{Gemmeke2017AudioSet}. In specialized audio domains, constructing such datasets is often prohibitively expensive and time-consuming, since annotation typically requires domain expertise, careful quality control, and repeated verification. Manual labeling may also introduce annotation bias and privacy risks, particularly when recordings contain sensitive acoustic information. For instance, medical audio analysis requires trained healthcare professionals to annotate heart sounds or respiratory signals~\cite{li2024matsed,liu2024taming}, while environmental sound analysis often relies on expert knowledge to identify specific acoustic events from long-duration recordings~\cite{piczak2016deep,mazzon2024synthio}. By contrast, unlabeled audio data are increasingly abundant across real-world scenarios. This imbalance between scarce annotations and abundant raw audio has made self-supervised learning (SSL) a compelling paradigm for audio representation learning, enabling models to exploit unlabeled data and improve transferability across downstream tasks. As shown in Fig.~\ref{fig:search}, annual publication counts suggest a rapid increase in research activity related to audio SSL.

Beyond the annotation bottleneck, supervised audio learning also faces limitations in generalization and robustness. Models trained with predefined labels may rely on dataset-specific shortcuts or spurious correlations rather than learning the underlying acoustic structures, leading to performance degradation under domain shifts, noisy recording conditions, or adversarial perturbations~\cite{goodfellow2014explaining,szegedy2013intriguing,zhang2016understanding}. SSL addresses these limitations by deriving supervisory signals from the intrinsic structure of audio signals rather than from manual annotations. In doing so, it encourages models to capture latent temporal, spectral, and contextual regularities from raw audio. Such supervision is typically instantiated through carefully designed \emph{proxy} or \emph{pretext tasks}, which define the learning objectives used to extract general-purpose acoustic representations from unlabeled audio.

Building on this principle, audio SSL methods differ primarily in how they formulate pretext tasks and derive training signals from unlabeled data. Predictive learning methods train models to infer future or missing representations from surrounding acoustic context~\cite{schneider2019wav2vec}, while masked modeling approaches corrupt portions of the raw waveform or time-frequency representation and require the model to recover the masked content or its latent targets~\cite{park2019specaugment}. Contrastive learning defines supervision by distinguishing related audio views from unrelated samples, often through objectives such as InfoNCE. Clustering-based methods, represented by frameworks such as HuBERT, generate discrete targets from the data itself and iteratively refine them to capture higher-level acoustic and semantic structures~\cite{baevski2020wav2vec2}. After pre-training, SSL models can be fine-tuned or used as feature extractors for diverse downstream applications, including speech recognition~\cite{baevski2020wav2vec2}, music genre classification~\cite{choi2017unsupervised}, and sound event detection~\cite{miao2026dynamic}. Compared with purely supervised models, SSL-based representations often provide stronger data efficiency and improved robustness in low-resource, noisy, and cross-domain conditions, making SSL a central direction in modern audio representation learning.

Recent progress further suggests that audio SSL is moving beyond task-specific representation learning toward unified audio understanding and native large audio language models (LALMs). Earlier SSL models were mainly used as general-purpose encoders for discriminative downstream tasks. More recent models aim to learn shared representations across heterogeneous audio domains, including speech, music, and environmental sounds. Representative models such as SPEAR~\cite{yang2025spear} and Dasheng~\cite{dinkel2024scaling} reflect this trend toward scalable and general-purpose audio understanding. In parallel, techniques such as multi-codebook vector quantization (MVQ) provide a promising mechanism for encoding audio at multiple granularities, from local acoustic details to higher-level event structures. More importantly, audio is increasingly becoming a core modality for generative modeling, instruction following, and multimodal reasoning, rather than merely a signal to be classified or recognized.

Despite these advances, existing surveys~\cite{hassani2020contrastive,liu2022audio} mainly provide foundations for contrastive learning, pretext task design, and SSL-based discriminative transfer. They do not fully capture recent developments in unified audio models, native LALMs, and safety-aware large-scale pre-training. As audio SSL continues to scale in data, model size, and downstream capabilities~\cite{deng2025scaling}, issues such as computational efficiency, deployment cost, privacy leakage, and safety alignment have become increasingly important. For example, studies on the HearSay benchmark~\cite{wang2026hearsay} show that advanced SSL representations may encode sensitive social attributes from voice signals even without explicit semantic cues. These emerging developments call for a timely and systematic reappraisal of audio SSL, which is the central goal of this survey.

\begin{figure}[!t]
    \centering
    \includegraphics[width=\linewidth]{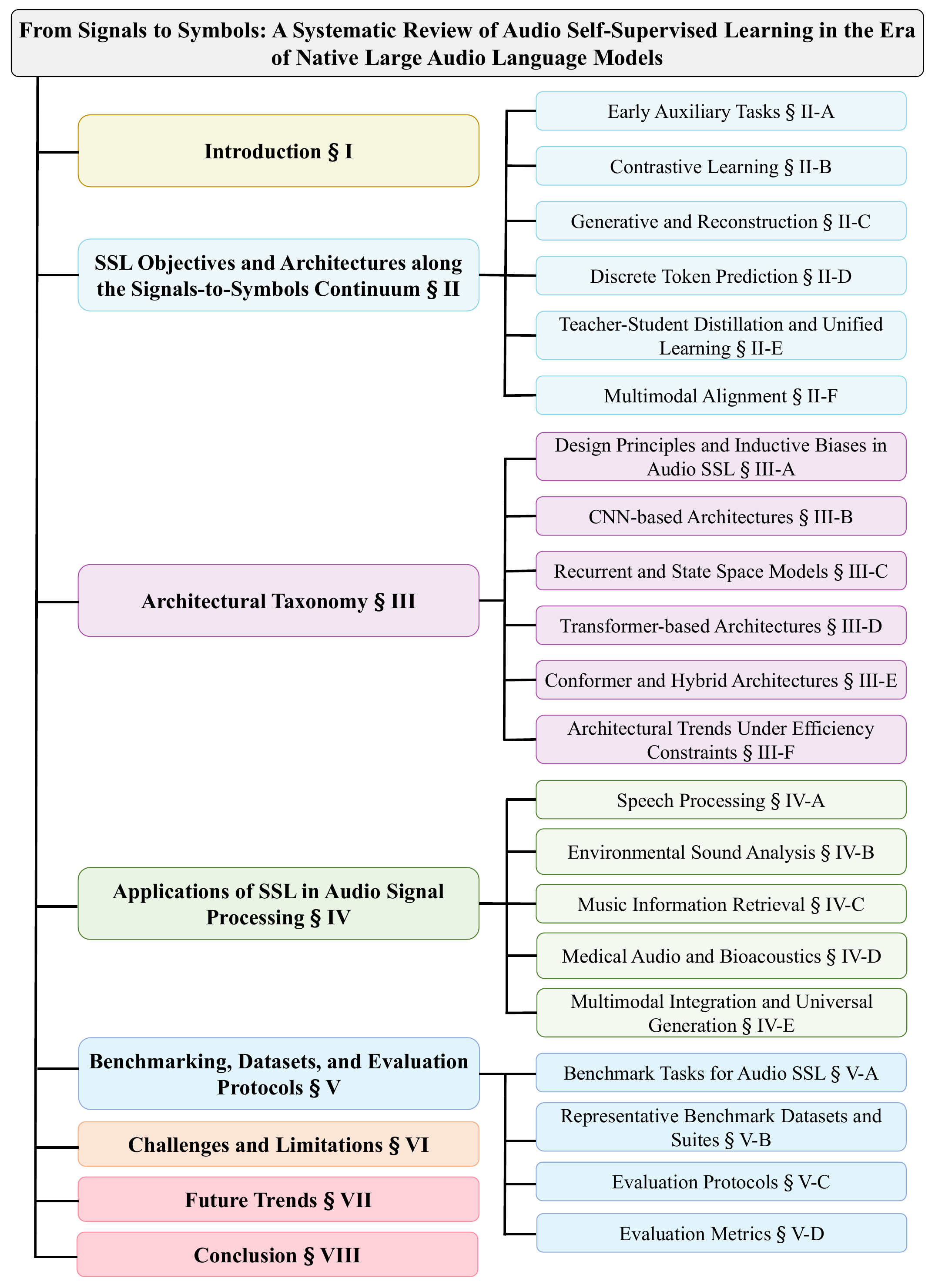}
    \caption{Framework of this paper.}
    \label{fig:framework}
\end{figure}

\smallskip
\noindent \textbf{Paper Organization.}
The remainder of this paper is organized as follows. Section~\ref{sec:taxonomy} presents a taxonomy of audio SSL methods, focusing on pretext tasks and learning objectives. Section~\ref{sec:Architectures} reviews input representations and backbone architectures, including waveform- and spectrogram-based models, CNNs, Transformers, and emerging SSM/Mamba designs. Section~\ref{sec:applications} summarizes applications in speech, music, and environmental sound analysis. Section~\ref{sec:benchmarking} discusses evaluation protocols and benchmarks, from SUPERB and HEAR to recent safety-oriented evaluations. Section~\ref{sec:challenges} analyzes key challenges, including privacy risks, computational efficiency, robustness, and polyphonic complexity. Section~\ref{sec:future_trends} outlines future directions toward general-purpose audio intelligence. Finally, Section~\ref{sec:conclusions} concludes the paper. Fig.~\ref{fig:framework} illustrates the overall framework of this survey.

\section{Objectives in Audio SSL: Representation Demands and Architectural Implications}
\label{sec:taxonomy}

This section examines audio self-supervised learning from the perspective of representation learning and model requirements. Instead of reviewing SSL methods as a chronological sequence of loss functions, we ask a more basic question: what kinds of representations are encouraged by different self-supervised objectives, and what do these objectives require from the underlying neural architecture?

Existing surveys have already provided comprehensive reviews of the main SSL pipelines, chronological developments, and representative models in audio and speech~\cite{8_ssl_survey, liu2022audio, Mohamed2022SelfSupervisedSR}. Our emphasis here is different. We treat the pretext objective as a main source of pressure on the learned representation. By defining which information should be preserved, suppressed, recovered, abstracted, or aligned, different objectives lead to different representation profiles. These differences then translate into different requirements for information flow, context aggregation, and temporal state propagation. In this sense, the objective is not only a training signal; it also shapes what the architecture needs to retain, combine, and infer.

This view is particularly useful for audio, where local acoustic patterns, long-range temporal dependencies, speaker and environmental variation, and semantic content are closely intertwined \cite{gong2022ssast, wu2018unsupervised}. As a result, objective design affects not only what the model learns, but also how evidence must move through the network. Some objectives depend mainly on local acoustic regularities, some require causal or sequential state updates, some rely on global inference from partial observations, and others require semantic correspondence across modalities.

Based on this view, we group existing objectives into five broad paradigms. Early auxiliary tasks derive supervision from manual transformations or temporal relations and usually encourage sensitivity to local structure and stable extraction of short-range regularities \cite{gidaris2018unsupervised, 87_jigsaw}. Contrastive learning uses agreement between augmented views and separation from mismatched instances, encouraging invariant representations while suppressing nuisance acoustic variation \cite{chen2020simple, saeed2021contrastive, baevski2020wav2vec2}. Generative and reconstruction objectives recover masked or corrupted inputs, which places more emphasis on contextual inference and information routing from sparse visible evidence \cite{MaskedAutoencoders2021, huang2022masked, chong2023masked}. Discrete token prediction uses clustered or quantized targets and shifts the learning goal from signal fidelity toward semantic abstraction and long-range reasoning over compressed states \cite{Guo2025RecentAI, chen2023beats, hsu2021hubert}. Multimodal alignment uses temporally co-occurring text or video and requires semantic projection and correspondence across heterogeneous modalities \cite{akbari2021vatt, laionclap2023, elizalde2023clap}.

\begin{figure*}[!t]
    \centering
    \includegraphics[width=0.85\linewidth]{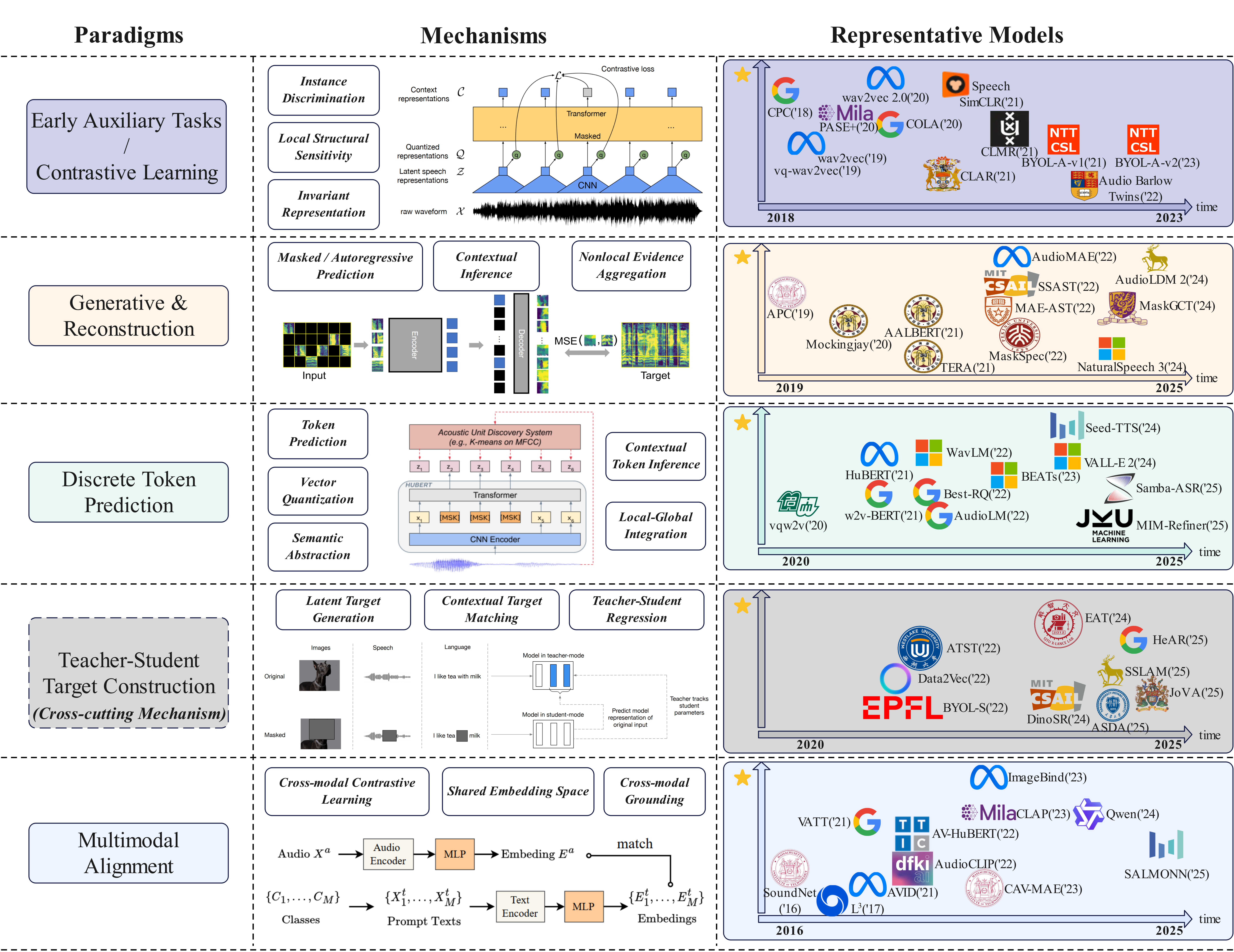}
    \caption{Objective-demand alignment in audio SSL. Beyond a chronological progression, the figure maps each supervisory paradigm to its characteristic processing demands and representation goals, ranging from local structural sensitivity and invariant representation learning to contextual inference, semantic abstraction, teacher-student target construction, and multimodal grounding. This mapping connects high-level SSL objectives with representative models and provides the basis for analyzing their architectural requirements in Section~\ref{sec:Architectures}.}
    \label{fig:trajectory}
\end{figure*}

As shown in Fig.~\ref{fig:trajectory} and summarized in Table~\ref{tab:ssl_audio_taxonomies}, these objective families differ not only in the form of supervision but also in the type of architectural support they need. The following subsections therefore do more than list representative methods. For each paradigm, we ask two related questions: what kind of representation does the objective encourage, and what architectural implications follow from that demand? This view also provides a bridge to Section~\ref{sec:Architectures}, where we examine why different supervisory signals tend to match different structural priors and inductive biases.

\subsection{Early Auxiliary Tasks}
\label{subsec:early_tasks}

Early work in audio SSL was largely based on manually designed pretext tasks that drew supervision from temporal order, local transformations, or cross-channel coincidence in unlabeled signals~\cite{8_ssl_survey}. Typical examples include temporal-gap prediction, as in Audio2Vec~\cite{tagliasacchi2020pre}, jigsaw-style reordering~\cite{87_jigsaw, misra2016shuffle, carr2021self}, and multi-task feature regression, as in PASE and PASE+~\cite{pascual2019learning, ravanelli2020multi}. The main contribution of this stage was not high-level semantic modeling itself, but the demonstration that unlabeled audio contains useful structural regularities.

In representation terms, these objectives mainly encourage sensitivity to local structure rather than semantic abstraction. Solving such heuristic tasks requires the model to preserve short-range acoustic continuity, spectral texture, and low-level physical cues such as pitch, energy, and formant transitions. The resulting representations therefore remain closely tied to local time-frequency structure and coarse relations among nearby segments.

The processing requirements of these objectives are correspondingly modest. They depend mostly on stable local evidence rather than long-range contextual inference. In most cases, the model needs to capture short-term temporal consistency, local ordering cues, or coarse correspondence across neighboring segments, rather than recover missing content from sparse observations or infer semantically meaningful units from distributed context. What matters most is reliable extraction of local acoustic primitives, while dynamic global routing plays a less central role.

This pattern also has a clear architectural consequence. Because the supervisory signal is grounded mainly in short-range regularities, these objectives fit naturally with architectures that impose a strong locality bias, especially convolutional encoders and related local front-end modules. The restricted receptive fields of such models are well suited to capturing stable low-level patterns from raw waveforms or time-frequency inputs without extensive global interaction.

The limits of this family are equally apparent. Because the supervisory signal is largely determined by task design rather than by the semantic structure of audio, the learned features often remain tied to local acoustic statistics or handcrafted heuristics, with limited abstraction and transferability across tasks~\cite{Mohamed2022SelfSupervisedSR}. As downstream objectives shifted toward long-range dependency modeling and higher-level semantic content, these early pretext tasks became less suitable. This limitation partly motivated the move toward contrastive learning and reconstruction-based paradigms.

\subsection{Contrastive Learning}
\label{subsec:contrastive}

Contrastive learning marks a shift from manually designed pretext tasks to supervision based on relations among views, time steps, or instances. By bringing matched samples closer and separating mismatched ones, or by aligning positive pairs under asymmetric objectives, these methods learn from relational structure in the data rather than from manually specified heuristics~\cite{wu2018unsupervised, chen2020simple}. In audio, representative examples include augmentation-based instance discrimination, such as COLA, CLAR, and BYOL-A~\cite{saeed2021contrastive, AlTahan2020, niizumi2023byol-a_v2}, predictive contrast over time, such as CPC and wav2vec~\cite{oord2018representation, schneider2019wav2vec}, and masked contextual contrast, as in wav2vec 2.0~\cite{baevski2020wav2vec2}. This transition mattered because it replaced task-specific heuristic supervision with a more general objective for learning transferable representations from unlabeled audio.

From a representation perspective, contrastive objectives mainly encourage invariance. To reduce the contrastive loss, the model must suppress nuisance variability, such as background noise, channel distortion, phase variation, or synthetic augmentation, while preserving information that remains stable across views or contexts. The goal is therefore not detailed reconstruction of the input, but learning a discriminative latent space in which semantically or structurally related samples stay close. The processing requirements are more flexible than in early auxiliary tasks, but they still depend strongly on how positive and negative pairs are defined. In augmentation-based contrastive learning, the main requirement is stable extraction and aggregation of acoustic cues that remain reliable under local perturbations. In predictive contrast, the model must also propagate temporal context in order to discriminate future or masked latent targets. Contrastive learning therefore combines two needs: robust local feature encoding and enough contextual aggregation to support discrimination across views, time steps, or instances.

These requirements have clear architectural implications. When the contrastive signal is driven mainly by local perturbations or waveform-level augmentations, architectures with a strong locality bias, especially convolutional encoders and related front-end modules, remain a good match because they support stable extraction of short-range acoustic structure~\cite{niizumi2021byol}. As the objective shifts toward predictive or masked contrast, however, local filters alone become less adequate because the model must also integrate broader temporal or contextual evidence. This helps explain the rise of sequential and hybrid architectures in contrastive audio SSL. Predictive coding objectives fit naturally with autoregressive structures that propagate temporal states, whereas masked contrastive frameworks such as wav2vec 2.0 are better served by a combination of local acoustic compression and global contextual modeling. The limits of this family arise from the same source as its strength. Because the learned invariances are strongly shaped by augmentation design or context sampling~\cite{8_ssl_survey}, it is often difficult in audio to decide which factors should be suppressed and which should be preserved, especially when speaker traits, prosody, recording conditions, and semantic content are tightly entangled. Under these conditions, contrastive learning may underemphasize fine-grained structure that is important for dense prediction, generation, or reconstruction-oriented tasks. This limitation partly motivated the later shift toward generative and reconstruction-based objectives.

\subsection{Generative and Reconstruction Objectives}
\label{subsec:generative}

Generative and reconstruction objectives start from a different assumption than contrastive learning. Instead of defining supervision through invariance across views, they ask the model to recover masked, corrupted, or future content from incomplete observations. Representative approaches include causal autoregressive prediction, as in APC~\cite{chung2019unsupervised}, bidirectional masked frame prediction, as in Mockingjay and Audio ALBERT~\cite{liu2020mockingjay, chi2021audio}, and patch-based masked autoencoding, as in Audio-MAE, SSAST, and MaskSpec~\cite{huang2022masked, gong2022ssast, chong2023masked}. More recent work has also connected reconstruction-based representation learning with neural synthesis, as in AudioLDM 2 and MaskGCT~\cite{liu2024audioldm2, wang2024maskgct}. Here, however, the main point is not the downstream generative application itself, but the fact that recovering acoustic targets can preserve rich contextual and structural information in the learned representation.

From a representation perspective, these objectives place more emphasis on signal fidelity and contextual continuity than strongly discrimination-based objectives. To recover missing spectrogram patches or future acoustic content, the model must retain both detailed local structure and broader context that helps infer what is missing. The resulting representation therefore tends to preserve richer acoustic information, which is often important for dense prediction, synthesis, and other tasks that depend on fine-grained structure. The processing demands also differ from those of contrastive learning. Reconstruction turns representation learning into a problem of contextual inference: the model must infer missing content from partial evidence rather than identify what remains invariant across views. This becomes especially clear under high masking ratios, such as the 80\% masking setting used in Audio-MAE~\cite{huang2022masked}, where visible evidence is sparse and distributed. Under these conditions, the architecture must aggregate information across distant temporal and spectral regions, which creates a strong need for dynamic information routing.

These demands have direct architectural implications. Because fixed local receptive fields are often not enough to bridge heavily masked gaps, masked reconstruction objectives align more naturally with architectures that support content-dependent global interaction. This helps explain the central role of Transformer-based architectures, where self-attention enables flexible long-range interaction, in patch-based masked audio modeling. At the same time, causal predictive reconstruction, as in APC, remains naturally compatible with recurrent or state-space models that propagate sequential context over time. The limitations of this family are also important. Because the objective requires reconstruction of continuous acoustic detail, the model may allocate substantial capacity to low-level variability, such as background noise, channel artifacts, or other non-semantic factors, that is not always useful for higher-level abstraction. This tension between signal fidelity and semantic condensation partly motivated the later rise of discrete token prediction objectives.

\subsection{Discrete Token Prediction}
\label{subsec:discrete_token}

Discrete token prediction grows out of a basic limitation of continuous reconstruction: not all recoverable acoustic detail is equally useful for representation learning. Rather than requiring the model to reproduce exact waveforms or spectrograms, this family asks it to predict clustered, quantized, or otherwise discrete targets derived from the input. Representative examples include offline clustering approaches such as HuBERT~\cite{hsu2021hubert}, models that combine masked prediction with contrastive learning such as w2v-BERT~\cite{chung2021w2vbert}, robustness-oriented extensions such as WavLM~\cite{chen2022wavlm}, and semantically informed tokenizers for general audio such as BEATs~\cite{chen2023beats}. The key shift is not quantization alone, but a change in the pretraining objective from signal fidelity to more structured categorical abstraction. The design and evaluation of discrete audio representations have recently become an active research topic in their own right~\cite{mousavi2025discrete}.

From a representation perspective, this objective encourages the model to suppress redundant acoustic variation while preserving higher-level phonetic, event-level, or semantic structure~\cite{Guo2025RecentAI}. Because the target is no longer the raw signal itself, the model is encouraged to map continuous acoustic input into more compact latent categories. The processing demands reflect a dual requirement: local acoustic compression and broader contextual inference. To predict a masked discrete token, the model must preserve enough local acoustic evidence to support target formation while also aggregating context to infer which latent category is most plausible in the masked region. In this sense, discrete token prediction shifts the object of inference from continuous signal recovery to semantically compressed target prediction, which creates a stronger need for dynamic information routing across compressed representations.

These demands have clear architectural implications. Discrete token prediction is most naturally supported by architectures that combine local acoustic encoding with stronger contextual modeling, rather than by purely local or purely reconstructive designs alone. This helps explain the prevalence of hybrid structures in this family, such as CNN-based front ends for acoustic compression followed by Transformer-style contextual encoders that support content-dependent global routing and masked token inference, as in HuBERT and WavLM~\cite{hsu2021hubert, chen2022wavlm}. More broadly, this paradigm favors architectures that can connect two levels of representation at once: low-level signal structure and higher-level categorical abstraction. Its limitations are equally important. The quality of the learned representation is strongly shaped by the tokenizer, clustering procedure, or teacher model used to construct the targets. In other words, discrete objectives do not remove inductive bias; they relocate part of it to the target-generation pipeline. If the discrete targets remain too closely tied to shallow acoustic variation, the resulting capacity for semantic abstraction is correspondingly limited. This boundary naturally motivates the next step toward multimodal alignment, where external semantic supervision is introduced more explicitly.

\subsection{Teacher-Student Target Construction}
\label{subsec:distillation}

As the preceding paradigms suggest, recent audio SSL methods increasingly blur the boundaries among contrastive, reconstructive, and discrete-target objectives. In this setting, teacher-student distillation is better viewed not as a separate objective family, but as a cross-cutting mechanism for constructing prediction targets. Representative frameworks such as data2vec~\cite{baevski2022data2vec} and efficiency-oriented designs such as EAT~\cite{chen2024eat} show that the student can be trained to match contextualized latent targets produced by a teacher network, rather than directly predict raw signals, negative pairs, or fixed cluster labels.

From a representation perspective, this mechanism shifts the source of supervision from manually specified targets to learned target structure. The prediction target is no longer defined only by physical fidelity, instance discrimination, or offline quantization. Instead, it is generated from a contextualized latent space. This suggests that representation quality depends not only on the loss function, but also on how the target is constructed, stabilized, and refined during training. Related ideas also appear in multi-objective settings such as PASE+~\cite{ravanelli2020multi}, where different supervisory signals are combined to encourage a shared encoder to capture more general-purpose acoustic structure.

From the perspective of processing demands, distillation does not introduce a completely separate architectural requirement in the same way as the objective families discussed above. What it changes is the form of the prediction target and, with it, the way existing architectures are used. In typical formulations, the teacher operates on less corrupted or more complete inputs and produces stable latent targets, whereas the student is trained on masked, perturbed, or otherwise degraded views. To bridge this information gap and match the teacher target, the student cannot rely on local acoustic evidence alone. It must aggregate enough context to infer the missing semantic structure. This creates a stronger need for contextual inference and turns the student task into one of dynamic information routing across sparse observations.

These asymmetric demands help explain why teacher-student distillation is often paired with architectures that combine stable acoustic encoding and strong contextual modeling. In practice, this often leads to locality-aware front ends together with Transformer-style encoders that support content-dependent global routing. The significance of distillation, however, lies less in defining a new architectural family than in showing that representation quality depends not only on what is predicted, but also on how the prediction target is generated and shared across tasks. The limits of this mechanism are also clear. Because the teacher is usually bootstrapped from the audio modality itself, the learned representation remains bounded by the semantic information available in audio alone. This limitation naturally motivates the next step toward multimodal alignment, where external semantic supervision is introduced more explicitly.

\subsection{Multimodal Alignment}
\label{subsec:multimodal}

Multimodal alignment extends self-supervised learning beyond the internal structure of the audio signal by using naturally co-occurring modalities, most notably language and vision, as external supervisory anchors. Representative frameworks include audio-visual correspondence models such as $L^3$, SoundNet, and AVID~\cite{arandjelovic2017look, SoundNet, morgado2021robust}, audio-language contrastive models such as CLAP and AudioCLIP~\cite{elizalde2023clap, laionclap2023, guzhov2022audioclip}, and unified multimodal architectures such as VATT~\cite{akbari2021vatt}. Rather than reconstructing masked inputs or predicting discrete acoustic tokens, this paradigm trains the model to align paired multimodal streams in a shared latent space, which makes it especially relevant for retrieval, transfer, and open-vocabulary understanding.

From a representation perspective, the main effect of this objective is semantic grounding. The encoder is no longer asked only to model acoustic regularities or suppress nuisance variation within audio itself. Instead, it must learn representations that remain meaningful when compared with paired text or visual signals. This encourages the audio embedding to capture higher-level concepts, such as named sound events or descriptive semantic content, that are often difficult to infer reliably from isolated audio alone. In this sense, multimodal alignment shifts representation learning from audio-internal organization toward cross-modal semantic consistency.

The processing demands also differ from those of audio-native SSL. To align audio with language or vision, the model must encode acoustic evidence and project it into a space in which semantically related signals from different modalities can be compared and associated. This creates a stronger requirement for cross-stream correspondence, semantic projection, and global contextual matching. It also introduces a need for cross-modal routing, because the model must abstract audio features into a representation that remains compatible with the target modality.

These demands have direct architectural implications. Multimodal alignment is often supported by architectures that combine strong audio encoding with flexible cross-modal interaction, such as dual-encoder contrastive frameworks, shared embedding models, and Transformer-based multimodal architectures. In practice, this is reflected in the frequent use of contextual audio encoders that support content-dependent matching across modalities, as in CLAP, AudioCLIP, and VATT~\cite{elizalde2023clap, laionclap2023, guzhov2022audioclip, akbari2021vatt}. The limitations of this paradigm are equally important. The learned representation is partly constrained by the granularity, availability, and bias of the paired modality, and alignment with text or vision does not automatically guarantee faithful modeling of audio-specific structure. For example, if a caption provides only coarse semantic content, the aligned audio embedding may underrepresent acoustic nuances that remain important for audio-native tasks. For this reason, multimodal alignment is best understood not as a replacement for audio-native SSL, but as a complementary route to broader semantic transfer and open-vocabulary generalization.

\begin{table*}[!t]
    \centering
    \caption{Objective--architecture alignment in representative audio SSL paradigms.}
    \label{tab:ssl_audio_taxonomies}
    \resizebox{\linewidth}{!}{
    \renewcommand{\arraystretch}{1.28}
    \setlength{\tabcolsep}{6pt}
    
    \begin{tabular}{
    >{\raggedright\arraybackslash}m{0.16\linewidth}
    >{\raggedright\arraybackslash}m{0.19\linewidth}
    >{\raggedright\arraybackslash}m{0.17\linewidth}
    >{\raggedright\arraybackslash}m{0.34\linewidth}
    >{\raggedright\arraybackslash}m{0.14\linewidth}}
    \toprule
    \textbf{SSL Paradigm} 
    & \textbf{Supervisory Target} 
    & \textbf{Main Processing Demand} 
    & \textbf{Architectural Alignment} 
    & \textbf{Representative Methods} \\
    \midrule
    
    Early auxiliary tasks
    & Relative position, temporal order, segment distance, or handcrafted proxy targets
    & Short-range acoustic sensitivity and local temporal structure
    & Locality-aware encoders, especially CNN-based front ends, are usually sufficient because the target is mainly defined by local continuity and ordering cues.
    & Audio2Vec, permutation-based methods, PASE/PASE+~\cite{tagliasacchi2020pre,carr2021self,pascual2019learning,ravanelli2020multi} \\
    
    \midrule
    
    Contrastive learning
    & Agreement between positive views and separation from negative samples or distractors
    & Invariance learning, nuisance suppression, and temporal/contextual discrimination
    & CNN front ends provide stable acoustic compression, while recurrent or Transformer encoders support context aggregation for predictive or masked contrastive objectives.
    & CPC, wav2vec, wav2vec 2.0, COLA, BYOL-A~\cite{oord2018representation,schneider2019wav2vec,baevski2020wav2vec2,saeed2021contrastive,niizumi2021byol} \\
    
    \midrule
    
    Predictive and reconstruction objectives
    & Future, masked, corrupted, or missing acoustic content
    & Contextual inference from partial observations and recovery of missing structure
    & Bidirectional Transformers, RNNs, or SSMs are well matched because they aggregate nonlocal evidence or propagate sequential states beyond fixed local filtering.
    & APC, Mockingjay, Audio ALBERT, Audio-MAE, SSAST, MaskSpec~\cite{chung2019unsupervised,liu2020mockingjay,chi2021audio,huang2022masked,gong2022ssast,chong2023masked} \\
    
    \midrule
    
    Discrete target prediction
    & Clustered, quantized, or tokenizer-derived acoustic units
    & Local acoustic encoding and contextual inference over compact semantic targets
    & Hybrid local--global designs are common: local front ends preserve acoustic detail, while contextual encoders infer masked or latent categorical targets.
    & HuBERT, w2v-BERT, WavLM, BEATs~\cite{hsu2021hubert,chung2021w2vbert,chen2022wavlm,chen2023beats} \\
    
    \midrule
    
    Teacher-student target learning
    & Contextualized latent targets produced by a teacher network
    & Target stabilization, latent-space prediction, and robust representation transfer
    & Student encoders typically require strong contextual modeling to match teacher targets from masked or corrupted inputs; the mechanism is often combined with Transformer or hybrid backbones.
    & data2vec, EAT, ATST~\cite{baevski2022data2vec,chen2024eat,li2024atst} \\
    
    \midrule
    
    Multimodal alignment
    & Paired audio-text, audio-visual, or cross-modal embedding targets
    & Semantic projection, cross-modal correspondence, and open-vocabulary matching
    & Dual encoders or multimodal contextual encoders project audio into a shared semantic space, supporting retrieval, zero-shot transfer, and cross-modal reasoning.
    & $L^3$, SoundNet, AVID, CLAP, AudioCLIP, VATT~\cite{arandjelovic2017look,SoundNet,morgado2021robust,elizalde2023clap,laionclap2023,guzhov2022audioclip,akbari2021vatt} \\
    
    \bottomrule
    \end{tabular}
    }
\end{table*}

\subsection{Summary and Architectural Implications}
\label{subsec:taxonomy_summary}

Taken together, the objective families reviewed in this section differ not only in the form of supervision, but also in the representation pressures and processing demands they impose. Across these paradigms, the development of audio SSL can be viewed as a shift from local structural sensitivity and augmentation-driven invariance to contextual inference, semantically compressed prediction, and cross-modal semantic grounding. Table~\ref{tab:ssl_audio_taxonomies} makes this relationship explicit by summarizing the taxonomy as an objective-architecture alignment matrix. Rather than merely listing representative methods, the matrix relates each SSL paradigm to its characteristic processing demands and the architectural inductive biases that most naturally support them. This perspective provides the basis for Section~\ref{sec:Architectures}, where we examine how different neural architectures, from locality-biased CNNs to Transformer-based contextual encoders and State Space Models, respond to these objective-induced demands.

\section{Architectures for Audio SSL: From Local Encoding to Global Context Modeling}
\label{sec:Architectures}
The preceding section reviewed audio SSL from the perspective of pretext tasks and learning objectives. This section turns to the architectural side and asks a complementary question: \emph{what types of neural structures are needed to support different SSL objectives?} In audio SSL, architectures are not merely interchangeable backbones. They determine how acoustic information is compressed, propagated, contextualized, and aligned with the supervisory signals constructed from unlabeled data.

Different SSL objectives impose different processing demands. Auxiliary and contrastive objectives often rely on stable local acoustic patterns and augmentation-invariant representations. Predictive objectives require sequential state propagation over time. Masked reconstruction and discrete token prediction require inference from incomplete observations, while multimodal alignment requires flexible semantic projection across modalities. These demands suggest that audio SSL architectures should be understood through their inductive biases: locality, sequential memory, global interaction, hybrid integration, and scalable computation.

Accordingly, this section reviews input representations and major backbone families used in audio SSL. We first discuss design principles and input formats, including raw waveforms and time-frequency representations. We then review CNN-based architectures for local acoustic encoding, recurrent and State Space Model (SSM)-based architectures for sequential modeling, Transformer-based architectures for global contextual interaction, and hybrid designs that combine local and global biases. Finally, we summarize efficiency-oriented trends that support scalable and deployable audio SSL. Fig.~\ref{fig:architect_overview} illustrates the overall landscape, and Table~\ref{tab:architecture_taxonomy} summarizes representative architectures.

\subsection{Design Principles and Input Representations}
\label{subsec:design_principles}
Audio signals differ from text and images in several important aspects. They evolve continuously over time, exhibit multi-scale temporal and spectral structures, and often contain causal or quasi-causal dynamics. Since SSL does not rely on explicit human labels, architectural inductive biases play a central role in determining which acoustic regularities can be extracted from unlabeled data~\cite{purwins2019deep}.

Two input formats dominate current audio SSL systems. The first is the raw waveform, which preserves the physical signal, including phase information. This representation is important for tasks such as speech enhancement, source separation, and high-fidelity generation, but it also poses a substantial dimensionality challenge: one second of audio sampled at 16 kHz contains 16,000 samples. Consequently, waveform-based SSL systems often rely on front-end encoders with strong locality assumptions to extract and compress stable low-level acoustic primitives before broader contextual reasoning becomes tractable. Representative systems such as wav2vec 2.0~\cite{baevski2020wav2vec2}, HuBERT~\cite{hsu2021hubert}, and WavLM~\cite{chen2022wavlm} use multi-layer 1D CNNs for this purpose, transforming raw samples into compact latent sequences for subsequent contextual modeling.

\begin{figure*}[!t]
    \centering
    \includegraphics[width=.8\linewidth]{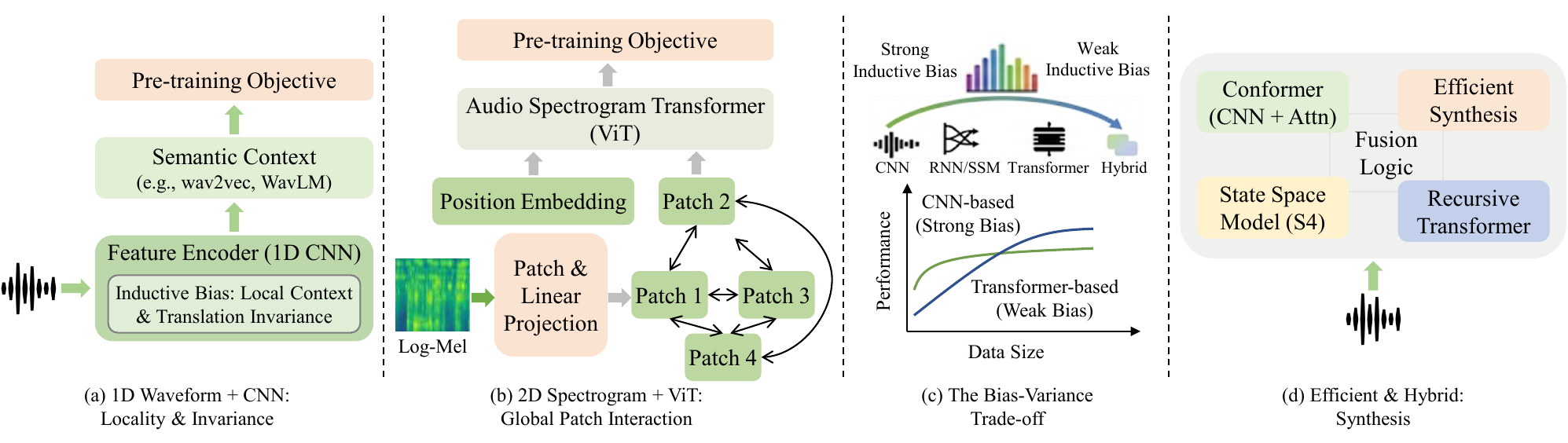}
    \caption{Landscape of objective-architecture alignment in audio SSL. It relates audio input formats and SSL objectives to representative architectural biases, including CNN-based locality, sequential state propagation, and Transformer-based global contextual interaction.}
    \label{fig:architect_overview}
\end{figure*}

The second widely used format is the time-frequency representation, such as the log-Mel spectrogram. Spectrograms provide a perceptually organized view of audio and are widely used in speech, music, and environmental sound analysis. However, unlike natural images, the two axes of a spectrogram have different meanings: the frequency axis captures pitch, formants, and timbre, whereas the time axis captures rhythm and temporal evolution. This anisotropic structure makes spectrogram modeling particularly suitable for masked reconstruction and discrete token prediction, where the model must infer missing or latent acoustic structures from visible time-frequency context. Under high masking ratios, as in SSAST~\cite{gong2022ssast} and AudioMAE~\cite{huang2022masked}, local continuity alone is often insufficient, and the architecture must support interaction across distant spectrogram regions. Fig.~\ref{fig:architect_overview} summarizes this objective-architecture alignment by relating audio input formats and SSL objectives to the corresponding architectural biases, including local acoustic encoding, sequential state propagation, and global contextual interaction.

These input characteristics lead to a broader trade-off among inductive bias, flexibility, and scalability. CNNs provide strong local priors and are effective for low-level acoustic encoding, but their fixed connectivity can limit nonlocal reasoning. Transformers offer flexible content-dependent interaction and are well aligned with masked modeling, discrete token prediction, and multimodal alignment, but their quadratic cost becomes expensive for long audio sequences. RNNs and SSM-based models provide an alternative for sequential state propagation and long-context modeling with more favorable scaling. Hybrid architectures combine these mechanisms to balance local acoustic grounding, global contextual reasoning, and computational efficiency. Table~\ref{tab:architecture_taxonomy} summarizes representative architectures under this objective-architecture alignment perspective.

\begin{figure*}[!t]
    \centering
    \includegraphics[width=.8\linewidth]{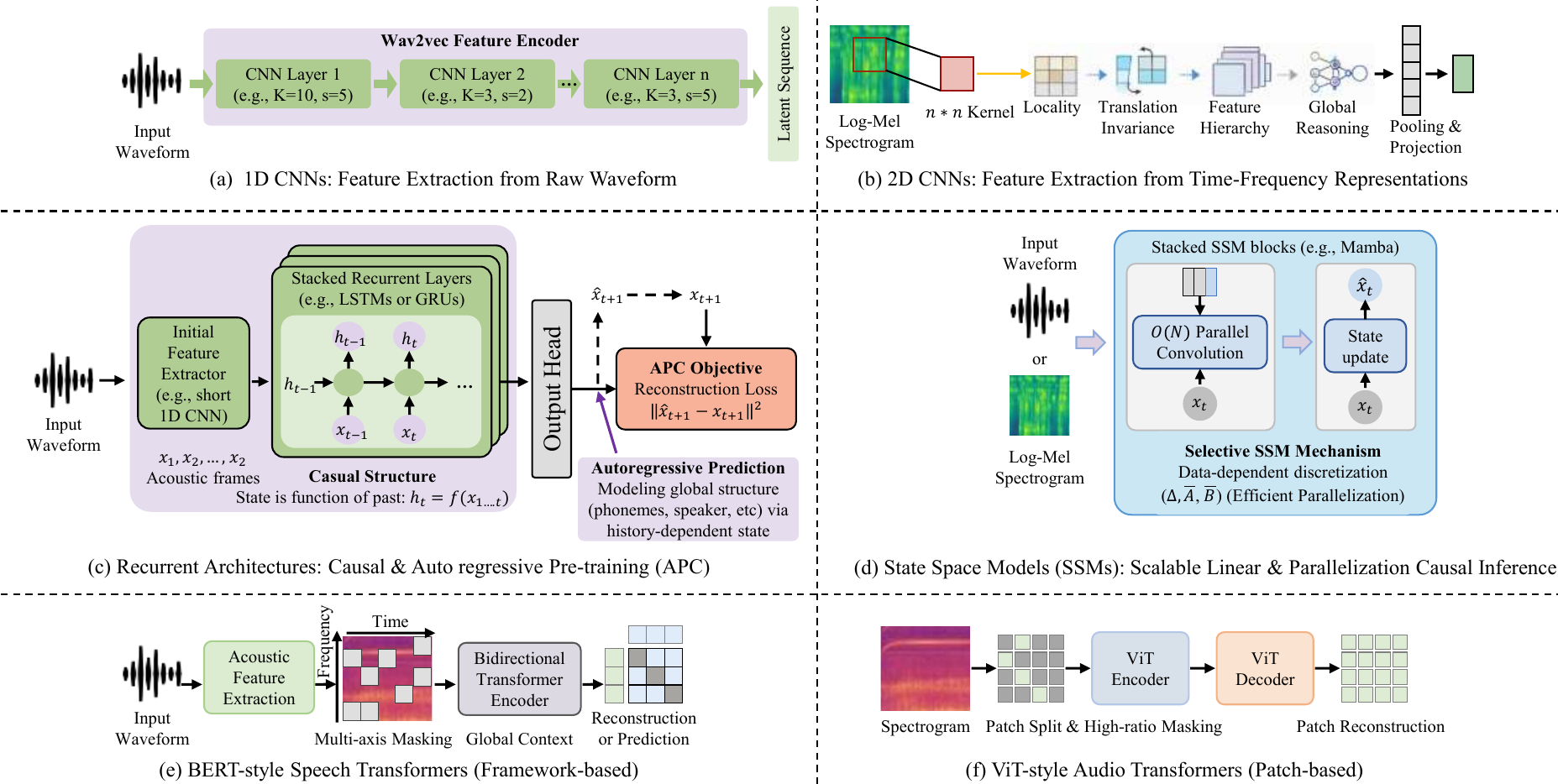}
    \caption{Representative structural mechanisms in audio SSL architectures. CNNs support local acoustic encoding, RNNs and SSM-based models support sequential state modeling, Transformers enable global contextual interaction, and hybrid architectures combine complementary inductive biases.}
    \label{fig:network_architect}
\end{figure*}

\begin{table*}[!ht]
\centering
\small
\caption{Objective-Architecture Alignment: Representative Audio SSL Backbones and Efficiency-Oriented Extensions.}
\renewcommand{\arraystretch}{1.18}
\resizebox{\linewidth}{!}{%
\begin{tabular}{p{2.3cm} p{4.2cm} p{4.2cm} p{3.0cm} p{1.6cm} p{3.6cm} p{2.0cm}}
\toprule
\textbf{Input Format} & \textbf{Architecture} & \textbf{Key Mechanism} & \textbf{Architectural Role} & \textbf{Scale} & \textbf{Representative Evidence / Role} & \textbf{Ref.} \\
\midrule

\rowcolor[gray]{0.92}
\multicolumn{7}{l}{\textbf{CNN-based Architectures}} \\

Raw waveform (1D)
& 1D CNN front end (wav2vec-style)
& Local convolution + strided downsampling
& Local acoustic compression
& 95M / 300M
& Canonical waveform SSL front end
& \cite{baevski2020wav2vec2} \\

Spectrogram (2D)
& EfficientNet in COLA
& Depthwise separable convolution
& Local time-frequency invariance
& 6M--66M
& Contrastive learning from local augmentations
& \cite{saeed2021contrastive} \\

\rowcolor[gray]{0.92}
\multicolumn{7}{l}{\textbf{Recurrent and State Space Models}} \\

Raw waveform
& LSTM / GRU in APC
& Gated recurrence with hidden state
& Causal state propagation
& $\sim$4M
& Autoregressive SSL baseline
& \cite{chung2019unsupervised} \\

Spectrogram
& Bidirectional Mamba in Audio Mamba
& Selective state space modeling
& Long-context state modeling
& $\sim$86M
& Alternative to attention-based contextual modeling
& \cite{yadav2024audiomamba} \\

Spectrogram
& Bidirectional Mamba in SSAMBA
& Selective SSM + bidirectional scan
& Scalable masked inference
& 13M--190M
& Reported ${\sim}95\%$ memory reduction in masked modeling
& \cite{shams2024ssamba} \\

Spectrogram
& BiMamba / ConExtBiMamba
& Selective SSM + external scan
& Bidirectional state propagation
& $\sim$20M--42M
& Efficient ASR-oriented sequence modeling
& \cite{zhang2025mamba} \\

Spectrogram
& xLSTM / AxLSTM
& Structured memory + enhanced gating
& Recurrent memory modeling
& $\sim$86M
& Modern recurrent-style long-sequence modeling
& \cite{beck2024xlstm, Yadav2024AxLSTMsLS} \\

\rowcolor[gray]{0.92}
\multicolumn{7}{l}{\textbf{Transformer-based Architectures}} \\

Spectrogram (frame)
& Mockingjay
& Bidirectional self-attention on frames
& Global contextual routing
& $\sim$30M
& Early BERT-style speech SSL
& \cite{liu2020mockingjay} \\

Spectrogram (frame)
& TERA
& Self-attention + multi-axis masking
& Robust masked contextualization
& $\sim$30M
& Robust representation under time-frequency corruption
& \cite{liu2021tera} \\

Spectrogram (patch)
& AST
& ViT patch embedding + self-attention
& Patch-level global interaction
& $\sim$87M
& Patch-based audio Transformer without a CNN front end
& \cite{gong2021ast} \\

Spectrogram (patch)
& SSAST
& Joint discriminative and generative patch objectives
& Global patch interaction
& $\sim$89M
& Self-supervised patch-based audio Transformer
& \cite{gong2022ssast} \\

Spectrogram (patch)
& Audio-MAE
& High-ratio masked autoencoding
& Nonlocal evidence aggregation
& $\sim$86M
& Masked spectrogram recovery from sparse patches
& \cite{huang2022masked} \\

Spectrogram (patch)
& BEATs
& Iterative acoustic tokenizer
& Semantic token prediction
& $\sim$90M
& Discrete acoustic target construction
& \cite{chen2023beats} \\

Spectrogram (patch)
& EAT
& Bootstrap learning with visible tokens
& Sparse global routing
& $\sim$86M
& Reported $15\times$ training speedup with visible-token processing
& \cite{chen2024eat} \\

Spectrogram (patch)
& ATST
& Dual-granularity teacher-student learning
& Hierarchical target construction
& $\sim$87M
& Clip-level and frame-level supervision
& \cite{li2024atst} \\

Spectrogram (patch)
& AaPE
& Aliasing-aware patch embedding
& Frequency-aware patch representation
& $\sim$86M
& Improved high-frequency representation
& \cite{yamamoto2025aape} \\

Waveform / spectrogram
& data2vec
& Contextual teacher target + masked prediction
& Unified target construction
& 95M / 317M
& Modality-general self-distillation
& \cite{baevski2022data2vec} \\

Spectrogram + text
& CLAP
& Dual-encoder contrastive alignment
& Cross-modal semantic alignment
& 123M--630M
& Audio-text representation and zero-shot transfer
& \cite{elizalde2023clap} \\

\rowcolor[gray]{0.92}
\multicolumn{7}{l}{\textbf{Conformer and Hybrid Architectures}} \\

Raw waveform
& CNN + Transformer in wav2vec 2.0
& CNN front end + Transformer encoder
& Local--global integration
& 95M / 300M
& Waveform compression with contextual prediction
& \cite{baevski2020wav2vec2} \\

Raw waveform
& CNN + Transformer in HuBERT
& CNN front end + clustered targets
& Local--global integration
& 95M / 317M
& Discrete prediction over latent units
& \cite{hsu2021hubert} \\

Raw waveform
& Conformer in WavLM
& Convolution module + self-attention
& Integrated local--global modeling
& 95M / 317M
& Strong transfer across speech tasks
& \cite{chen2022wavlm} \\

Raw waveform
& Conformer in BigSSL
& Large-scale Conformer encoder
& Scaled hybrid modeling
& 600M--1B+
& Foundation-scale speech representation learning
& \cite{zhang2022bigssl} \\

Raw waveform
& HyperConformer
& Lightweight mixing in hybrid blocks
& Efficient local--global fusion
& $\sim$30M
& Efficient hybrid contextual modeling
& \cite{mai2023hyper} \\

Spectrogram
& Branchformer / E-Branchformer
& Parallel attention and gated MLP branches
& Parallel local--global modeling
& $\sim$116M
& Decoupled local and global branches
& \cite{peng2022branchformer, kim2023ebranchformer} \\

Spectrogram
& Zipformer
& Variable frame-rate U-Net encoder
& Multi-scale temporal modeling
& 23M--148M
& Efficient multi-scale sequence encoding
& \cite{yao2024zipformer} \\

Spectrogram
& HTS-AT
& Window attention + token aggregation
& Hierarchical spectrogram modeling
& $\sim$31M
& Efficient long-form audio representation
& \cite{chen2022htsat} \\

\rowcolor[gray]{0.92}
\multicolumn{7}{l}{\textbf{Efficiency-Oriented Extensions}} \\

Spectrogram
& Joint-KD
& Semantic KD + masked acoustic modeling
& Distillation-based compression
& $\sim$30M
& Compact full-band speech representation under teacher guidance
& \cite{liu2024joint} \\

Raw waveform
& DistilHuBERT / SKILL
& Correlation or relational alignment
& Student representation compression
& 24M--26M
& Reported $75\%$ parameter reduction with preserved representation quality
& \cite{chang2022distilhubert, Zampierin2024SkillSK} \\

Spectrogram
& SemantiCodec
& Semantic discrete encoding
& Codec-based semantic compression
& $\sim$100M
& Low-bitrate semantic codec representation
& \cite{liu2024semanticodec} \\

Spectrogram
& WavTokenizer
& Single-codebook acoustic tokenizer
& Compact acoustic tokenization
& $\sim$37M
& Compact tokenizer for audio language modeling
& \cite{ji2025wavtokenizer} \\

Raw waveform
& EnCodec / DAC
& RVQ + encoder-decoder compression
& Neural audio compression
& 15M / $\sim$74M
& High-fidelity codec-style representation
& \cite{li2024mert, du2025codecfake} \\

Raw waveform
& TS3-Codec
& Causal masking + streaming design
& Streaming-compatible compression
& $\sim$40M
& Causal codec representation for streaming
& \cite{wu2024ts3} \\

Raw waveform
& Fast-HuBERT
& Blockwise downsampling encoder
& Efficient contextual inference
& $\sim$95M
& High-throughput SSL inference
& \cite{yang2023fasthubert} \\

\bottomrule
\end{tabular}
}
\label{tab:architecture_taxonomy}
\end{table*}

\subsection{CNNs: Local Acoustic Encoding}
\label{subsec:cnn_architectures}

CNNs provide one of the most established architectural priors in audio representation learning. In audio SSL, their importance does not only come from historical success in supervised acoustic modeling, but also from their compatibility with objectives that depend on local acoustic regularities and augmentation-invariant representations. Through local filtering, weight sharing, pooling, and strided downsampling, CNNs efficiently capture short-range temporal or time-frequency structures, such as phonetic onsets, formant transitions, harmonic patterns, and transient events~\cite{hershey2017cnn}. These properties make CNNs particularly suitable for early auxiliary tasks, augmentation-driven contrastive learning, and front-end acoustic compression.

CNNs are commonly used in two forms in audio SSL. The first is the 1D convolutional front end for raw waveform modeling. Since raw waveforms have high temporal resolution, direct contextual modeling over sample-level sequences is computationally expensive. Systems such as wav2vec 2.0~\cite{baevski2020wav2vec2}, HuBERT~\cite{hsu2021hubert}, and WavLM~\cite{chen2022wavlm} therefore use stacked temporal convolutions to transform raw samples into lower-rate latent sequences before applying higher-level contextual encoders. In this setting, the CNN front end serves both as a compression module and as an acoustic prior: it determines which low-level cues are preserved for subsequent contrastive prediction, masked modeling, or discrete target prediction.

The second form is the 2D convolutional backbone for spectrogram-based learning. Log-Mel spectrograms exhibit local time-frequency structures that can be effectively modeled by convolutional kernels. Contrastive and self-distillation methods such as COLA~\cite{saeed2021contrastive} and BYOL-A~\cite{niizumi2021byol} use CNN backbones, including ResNet or EfficientNet variants, to learn representations that remain stable under audio augmentations such as cropping, mixup, time-frequency masking, and local perturbations. In these cases, the main objective is to learn robust acoustic invariances from local evidence rather than to perform broad semantic reasoning over long contexts.

Despite these advantages, pure CNN architectures have clear limitations for recent audio SSL objectives. Masked reconstruction, discrete token prediction, teacher-student target learning, and multimodal alignment often require the model to infer missing or semantic information from distributed context. Although deeper, dilated, or multi-scale convolutions can enlarge the receptive field, the resulting interaction pattern remains largely fixed and locally biased. This makes CNNs less flexible than attention-based or state-space architectures when the objective requires content-dependent interaction across distant time-frequency regions or modalities.

Therefore, CNNs are efficient local encoders and acoustic compression modules rather than universal backbones for audio SSL. They remain highly useful when the objective emphasizes local invariance, waveform compression, or stable low-level feature extraction. For objectives requiring long-range contextual inference or semantic alignment, CNNs are often combined with Transformers, Conformers, recurrent models, or SSM-based architectures to form hybrid systems that integrate local acoustic grounding with broader contextual modeling.

\subsection{Recurrent and SSM Models: Sequential State Modeling}
\label{subsec:recurrent_ssm}

While CNNs emphasize local acoustic encoding and Transformers support global token interaction, another important line of audio SSL architectures focuses on sequential state modeling. This family is particularly relevant to objectives that require temporal propagation, causal prediction, streaming inference, or long-context acoustic modeling. In these settings, the model must maintain an informative state over time rather than treating each acoustic segment as an isolated pattern. Such a requirement naturally arises in predictive coding, autoregressive learning, and long-form audio representation learning, where phonetic transitions, prosody, speaker traits, and acoustic events unfold progressively along the temporal axis.

Early audio SSL methods were closely related to recurrent neural networks. Autoregressive Predictive Coding (APC)~\cite{chung2019unsupervised}, for example, uses LSTM- or GRU-based encoders~\cite{hochreiter1997long} to predict future acoustic frames from past observations. This design directly matches the causal nature of the pretext task: the recurrent hidden state summarizes previous acoustic evidence and provides a compact context for future prediction. Similar recurrent structures were also adopted in early speech representation learning because they provide an intuitive mechanism for modeling temporal continuity and sequential dependency. However, conventional RNNs are difficult to scale to modern SSL settings. Their sequential computation limits parallel training, and their effective memory becomes insufficient when modeling long audio streams or dense frame-level representations.

State Space Models (SSMs) have recently emerged as a promising alternative for sequential audio modeling. Unlike self-attention, which explicitly computes pairwise token interactions, SSMs propagate information through structured latent states with more favorable sequence-length scaling. Selective SSMs such as Mamba~\cite{gu2023mamba} further improve this paradigm by allowing input-dependent state updates, making them more flexible for complex acoustic sequences. From the perspective of audio SSL, SSMs are attractive because they offer a compromise between recurrent state propagation and scalable long-context modeling. They are especially relevant when the objective requires temporally coherent context but does not necessarily require full global pairwise attention.

Recent studies have begun to incorporate SSM-style backbones into audio SSL and speech representation learning. Audio Mamba~\cite{yadav2024audiomamba} and SSAMBA~\cite{shams2024ssamba} introduce bidirectional Mamba layers for audio representation learning, often under masked spectrogram modeling settings. These models suggest that masked inference and long-context acoustic modeling can be supported not only by Transformer-based self-attention, but also by scalable state-space sequence modeling. Related recurrent-style developments, such as xLSTM~\cite{beck2024xlstm} and AxLSTM~\cite{Yadav2024AxLSTMsLS}, also revisit memory-based sequence modeling through improved gating and memory mechanisms. Together, these methods indicate a renewed interest in sequential architectures for audio SSL, especially under efficiency and long-context constraints.

Overall, recurrent and SSM-based architectures occupy a complementary position in the audio SSL landscape. Compared with CNNs, they provide stronger temporal state propagation; compared with standard Transformers, they offer more favorable scaling for long sequences. Their main limitation is that they may provide less flexible content-dependent interaction than full self-attention, particularly for objectives requiring complex nonlocal association or multimodal alignment. Therefore, sequential models are best viewed as efficient long-context backbones for SSL objectives where temporal continuity, streaming capability, and scalable state modeling are central.

\subsection{Transformers: Global Context Modeling}
\label{subsec:transformer_arch}

Transformers have become one of the dominant architectures in audio SSL because many modern pretext tasks require flexible contextual modeling beyond local acoustic continuity. Compared with CNNs, which emphasize local time-frequency patterns, and recurrent or SSM-based models, which propagate information through sequential states, Transformers provide content-dependent interactions among frames, latent units, or spectrogram patches through self-attention~\cite{vaswani2017attention}. This property is particularly useful for masked modeling, discrete token prediction, teacher-student target learning, and multimodal alignment, where the model must infer missing or semantic structure from distributed acoustic evidence.

Early Transformer-based audio SSL methods were largely inspired by BERT-style masked prediction. Mockingjay~\cite{liu2020mockingjay} introduced bidirectional Transformer encoders for masked acoustic modeling, while TERA~\cite{liu2021tera} extended this idea by applying reconstruction objectives under time, frequency, and magnitude perturbations. These models demonstrated that bidirectional self-attention can learn contextual speech representations by integrating information from both nearby and distant frames. In contrast to autoregressive prediction, such bidirectional formulations are better suited to objectives where the target is inferred from incomplete but non-causal acoustic context.

The use of Transformers has further expanded with patch-based spectrogram modeling. Following the success of Vision Transformers in visual recognition~\cite{dosovitskiy2020image}, AST~\cite{gong2021ast} introduced patch-based audio Transformers for spectrogram inputs, providing an important basis for later SSL methods. SSAST~\cite{gong2022ssast} combined discriminative and generative objectives over spectrogram patches, while Audio-MAE~\cite{huang2022masked} adopted high-ratio masked autoencoding to learn representations from sparse visible patches. These methods show that self-attention is well suited to masked spectrogram learning, where the model must aggregate nonlocal evidence across the time-frequency plane rather than relying only on local continuity.

Recent Transformer-based SSL methods have also moved from low-level reconstruction toward semantic target prediction and more efficient pretraining. BEATs~\cite{chen2023beats} learns acoustic tokenizers and predicts discrete acoustic labels, thereby shifting the objective from waveform or spectrogram recovery to semantically richer representation learning. EAT~\cite{chen2024eat} improves the efficiency of masked audio pretraining by processing visible tokens, reducing unnecessary computation over masked regions. Other methods, such as ATST~\cite{li2024atst} and AaPE~\cite{yamamoto2025aape}, refine the design of teacher-student targets, temporal granularity, or patch embeddings. Together, these studies indicate that Transformer-based audio SSL has evolved from generic contextual encoders toward more structured and efficient pretraining frameworks.

Despite their effectiveness, Transformers are not a universal solution for audio SSL. Standard self-attention has quadratic complexity with sequence length, which becomes costly for long recordings, dense spectrogram patches, and audio-language modeling pipelines. Moreover, Transformers provide weaker built-in local acoustic priors than convolutional front ends, often requiring large-scale data or hybrid designs to learn robust low-level structures. These limitations explain why many successful systems combine Transformers with CNN front ends, Conformer blocks, efficient attention, token pruning, or SSM-based modules. Therefore, Transformers play a central role in audio SSL by enabling flexible global context modeling, particularly for objectives involving incomplete observations, semantic token prediction, or cross-modal alignment. At the same time, they remain complementary to architectures with stronger locality or sequence-efficiency biases.

\subsection{Hybrid Architectures: Local-Global Integration}
\label{subsec:conformer_hybrid}
\begin{figure*}[!t]
    \centering
    \includegraphics[width=.8\linewidth]{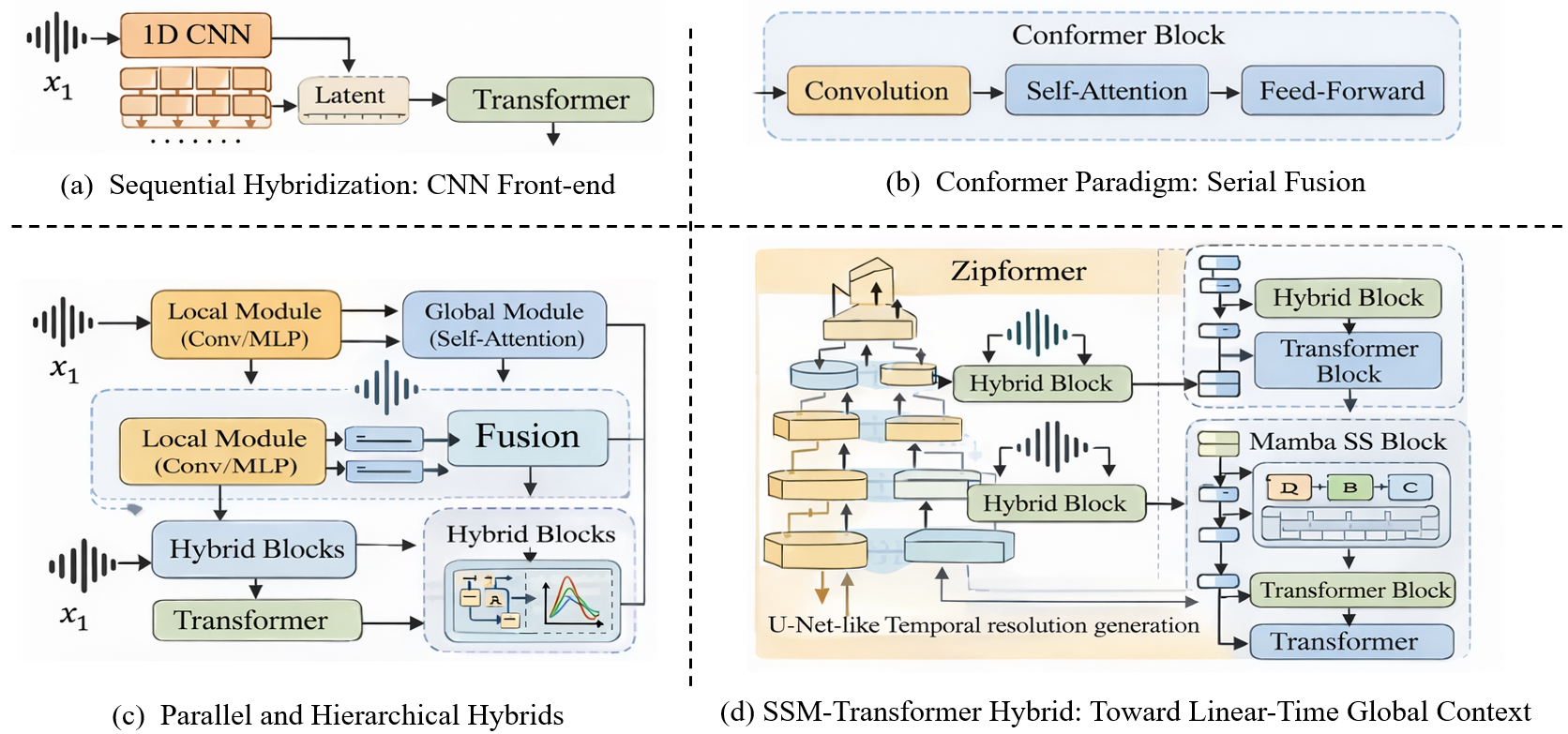}
    \caption{Hybrid architectures for local-global integration in audio SSL. Hybrid designs combine local acoustic encoding, content-dependent global interaction, and efficient long-context modeling. Sequential pipelines and Conformer-style blocks integrate convolutional modules with self-attention, while parallel, hierarchical, and SSM-augmented designs improve multi-scale representation learning and scalability.}
    \label{fig:architect_hybrid}
\end{figure*}

CNNs, sequential models, SSMs, and Transformers provide complementary inductive biases for audio SSL. CNNs capture local acoustic patterns, recurrent and SSM-based models propagate temporal states, and Transformers support global contextual interaction. Since many SSL objectives require these capabilities jointly, hybrid architectures have become a common design choice. As summarized in Fig.~\ref{fig:architect_hybrid}, hybrid architectures in audio SSL can be broadly grouped into sequential pipelines, Conformer-style integrated blocks, parallel and hierarchical hybrids, and SSM-augmented designs.

\emph{Sequential hybridization} is the most widely used pattern, typically combining a local front end with a contextual encoder. In waveform-based SSL, wav2vec 2.0~\cite{baevski2020wav2vec2}, HuBERT~\cite{hsu2021hubert}, and WavLM~\cite{chen2022wavlm} first use stacked temporal convolutions to compress raw audio into latent sequences, followed by Transformer or Conformer encoders for contextual prediction and discrete token inference. This design separates low-level acoustic encoding from high-level contextual modeling and has become a standard architecture in speech SSL.

\emph{Conformer-style integrated blocks} provide a tighter form of local--global fusion. The Conformer~\cite{gulati2020conformer} combines convolutional modules and multi-head self-attention within each block: the convolutional component preserves local continuity and short-range acoustic patterns, while self-attention captures longer-range dependencies. This balance is particularly useful for SSL objectives involving masking, denoising, contrastive learning, or teacher-student prediction, where both local robustness and global context are required.

\emph{Parallel and hierarchical hybrids} further extend local--global integration. Branchformer and E-Branchformer~\cite{peng2022branchformer,kim2023ebranchformer} use separate branches to model local and global dependencies before combining them, while Zipformer~\cite{yao2024zipformer} and HTS-AT~\cite{chen2022htsat} introduce multi-scale processing to handle long audio sequences. These designs are particularly relevant because audio events occur at multiple temporal scales, from short transients and phonetic cues to longer rhythmic, semantic, or environmental structures.

\emph{SSM-Transformer hybrids} are emerging as an efficiency-oriented form of hybridization. In these designs, attention provides content-dependent interaction, while SSM or Mamba-style modules support efficient long-context state modeling. This combination is increasingly relevant as audio SSL moves toward longer recordings, dense token sequences, and audio-language modeling pipelines, where full self-attention can become computationally expensive.

Overall, hybrid architectures reflect the fact that modern audio SSL objectives rarely depend on a single type of information processing. They combine local acoustic encoding, global contextual reasoning, multi-scale temporal modeling, and efficient sequence processing. Their flexibility makes them effective across diverse SSL settings, but also increases design complexity, requiring careful choices according to input format, pretext task, sequence length, and computational budget.

\subsection{Efficient and Deployable Architectures}
\label{subsec:efficiency_trends}

As audio SSL scales to larger datasets, longer sequences, and broader downstream applications, efficiency has become an increasingly important factor in architecture design. Computational cost, memory usage, inference latency, and sequence-length scalability directly affect whether SSL models can be trained, transferred, and deployed in practical scenarios. This issue becomes more pronounced for high-resolution waveforms, dense spectrogram patches, long token sequences, and audio-language modeling pipelines, where the cost of contextual modeling can grow rapidly. Therefore, recent audio SSL research has increasingly considered efficiency not only as an implementation detail, but also as a key design consideration that shapes model architecture, input representation, tokenization strategy, and adaptation method.

A major line of work focuses on compressing pretrained SSL models through knowledge distillation. Methods such as DistilHuBERT~\cite{chang2022distilhubert}, Joint-KD~\cite{liu2024joint}, and SKILL~\cite{Zampierin2024SkillSK} train compact student models to preserve the representational structure or semantic information learned by larger teacher models. These methods are motivated by the observation that large SSL encoders often contain rich contextual knowledge, but their parameter size and inference cost can limit their use in downstream systems. Distillation therefore provides a practical way to retain much of the benefit of large-scale pretraining while reducing memory and computational requirements.

Another important direction is representation-level compression through neural codecs and discrete tokenizers. Methods such as SemantiCodec~\cite{liu2024semanticodec} and WavTokenizer~\cite{ji2025wavtokenizer} convert continuous audio signals into compact discrete or codec-like units. These representations reduce sequence redundancy and provide structured targets for discrete prediction objectives. They are also increasingly relevant to audio language modeling, where tokenized audio units serve as the interface between acoustic encoders and generative language-model backbones. In this context, tokenization is not merely a preprocessing step; it affects the granularity, semantic content, and efficiency of subsequent SSL or generative modeling.

Efficiency constraints also affect model adaptation and real-time deployment. Parameter-efficient fine-tuning methods allow large SSL encoders to be adapted to downstream tasks without updating all parameters, reducing storage and optimization costs. Meanwhile, streaming and low-latency applications require models to produce representations incrementally with limited or no access to future context. This requirement favors designs based on causal masking, blockwise processing, progressive downsampling, sparse computation, or multi-scale encoding. Systems such as TS3-Codec~\cite{wu2024ts3}, Fast-HuBERT~\cite{yang2023fasthubert}, and Zipformer~\cite{yao2024zipformer} illustrate how deployment constraints can reshape context aggregation and sequence modeling strategies.

Overall, efficiency-oriented research broadens the architectural landscape of audio SSL beyond standard backbone selection. Model compression, tokenization, efficient sequence modeling, parameter-efficient adaptation, and streaming-compatible design all address the same underlying tension: how to preserve representation quality while reducing computational and deployment costs. As audio SSL moves toward unified audio understanding and large audio language models, this trade-off will become increasingly central to both architectural design and practical system deployment.

\begin{table*}[t]
\centering
\small
\caption{Task-mechanism alignment of SSL and SSL-derived representations in audio applications.}
\renewcommand{\arraystretch}{1.18}
\resizebox{\linewidth}{!}{%
\begin{tabular}{p{2.7cm} p{4.0cm} p{4.2cm} p{4.5cm} p{3.8cm}}
\toprule
\textbf{Application Area} & \textbf{Task-Specific Demand} & \textbf{Relevant SSL Mechanism} & \textbf{Representative Methods} & \textbf{Typical Evaluation} \\
\midrule

Speech processing
& Phonetic continuity, speaker/prosodic cues, multilingual and low-resource transfer
& Predictive coding, masked modeling, contrastive learning, discrete unit prediction
& wav2vec 2.0~\cite{baevski2020wav2vec2}, HuBERT~\cite{hsu2021hubert}, WavLM~\cite{chen2022wavlm}, XLS-R~\cite{babu2021xlsr}, SSL-AASIST~\cite{yamaguchi2025investigating}
& WER/CER, PER, EER, accuracy, F1 \\

Environmental sound analysis
& Polyphony, open event vocabulary, device mismatch, domain shift, anomaly detection
& Masked spectrogram modeling, acoustic token prediction, contrastive alignment, reconstruction/consistency learning
& Audio-MAE~\cite{huang2022masked}, BEATs~\cite{chen2023beats}, Dasheng~\cite{dinkel2024scaling}, CLAP~\cite{elizalde2023clap}, SSLAM~\cite{alex2025sslam}, Anopatch~\cite{jiang2024anopatch}, DP-MAE~\cite{liu2024dp}
& mAP, accuracy, event-F1, ER, localization error, AUC \\

Music information retrieval
& Polyphony, harmonic/rhythmic structure, timbre, long-range form, subjective labels
& Music-specific tokenization, masked modeling, audio-language alignment, mixture consistency
& MERT~\cite{li2024mert}, MuQ~\cite{zhu2025muq}, MusicFM~\cite{won2024foundation}, LAION-CLAP~\cite{laionclap2023}, MixIT~\cite{wisdom2020unsupervised}, Pac-HuBERT~\cite{chen2023pachubert}
& Accuracy, mAP, retrieval recall, SDR, structure metrics \\

Medical audio and bioacoustics
& Expert-label scarcity, physiological variability, cross-species transfer, rare events
& Masked modeling, contrastive learning, multimodal SSL, few-shot transfer
& HeAR~\cite{Baur2024HeARH}, SleepFM~\cite{thapa2024sleepfm}, DEPA~\cite{zhang2021depa}, AVES~\cite{hagiwara2023aves}
& AUROC, AUPRC, F1, accuracy, task-specific clinical/ecological metrics \\

Multimodal and generative audio
& Semantic grounding, audio-visual correspondence, audio-language reasoning, controllable generation
& Cross-modal contrastive learning, masked multimodal modeling, semantic conditioning, discrete audio tokenization
& CAV-MAE~\cite{gong2023contrastive}, MAViL~\cite{huang2023mavil}, ImageBind~\cite{girdhar2023imagebind}, SALMONN~\cite{tang2024salmonn}, Qwen2-Audio~\cite{Qwen2-Audio}, AudioLDM 2~\cite{liu2024audioldm2}
& Retrieval recall, mAP, accuracy, WER, captioning metrics, MOS, FAD \\

\bottomrule
\end{tabular}
}
\label{tab:task_mechanism_alignment}
\end{table*}

\section{Applications of SSL in Audio Signal Processing}
\label{sec:applications}

Audio SSL has become a general representation learning paradigm across a wide range of audio signal processing tasks. Its main value lies in reducing dependence on task-specific annotations, improving transferability across domains, and enabling robust modeling under low-resource, noisy, or distribution-shifted conditions. Building on the objective and architectural taxonomies discussed in Sections~\ref{sec:taxonomy} and~\ref{sec:Architectures}, this section reviews how SSL representations are used in five major application areas: speech processing, environmental sound analysis, music information retrieval, medical and biological acoustics, and multimodal audio understanding and generation. Table~\ref{tab:task_mechanism_alignment} summarizes representative applications, models, datasets, and evaluation protocols.

\subsection{Speech Processing}
\label{subsec:speech_apps}

Speech processing is the most established application area of audio SSL. Speech signals exhibit strong temporal continuity, hierarchical phonetic structure, and rich paralinguistic variation. These properties match the assumptions of predictive coding, masked modeling, and discrete unit prediction, where models learn to infer missing, future, or latent speech representations from surrounding acoustic context. Consequently, SSL encoders can acquire phonetic, speaker, and prosodic representations from unlabeled speech before adaptation to downstream tasks.

In automatic speech recognition (ASR), SSL is particularly effective in low-resource and multilingual settings. Models such as wav2vec 2.0, HuBERT, WavLM, and XLS-R learn transferable speech representations from large-scale unlabeled corpora, reducing the reliance on manually transcribed data during fine-tuning~\cite{baevski2020wav2vec2,hsu2021hubert,chen2022wavlm,babu2021xlsr}. Their effectiveness stems from the close match between SSL objectives and speech structure: masked or contrastive prediction encourages the encoder to recover phonetic and contextual information that is directly relevant to recognition. Benchmarks such as SUPERB further show that SSL representations support a broad set of speech tasks, including phoneme recognition, speaker identification, and emotion recognition~\cite{yang2021superb}.

Beyond ASR, SSL representations are widely used for paralinguistic analysis because they preserve non-lexical cues such as pitch, rhythm, speaking rate, speaker identity, and emotional prosody. These cues are difficult to annotate exhaustively but are naturally embedded in raw speech. This makes SSL backbones such as HuBERT, WavLM, and XLS-R useful for speech emotion recognition, speaker analysis, and related tasks~\cite{lian2025mer}. However, the same representational richness also raises privacy concerns, since speaker traits and sensitive attributes may be retained in pretrained embeddings.

Speech security provides another important application scenario. Synthetic speech and deepfakes often contain subtle artifacts caused by vocoders, neural codecs, or generation pipelines. These artifacts may appear in local spectral details, phase patterns, or long-range temporal dependencies. SSL encoders pretrained on large-scale natural speech provide multi-level representations that can expose such deviations more effectively than hand-crafted acoustic features. Recent anti-spoofing systems therefore exploit SSL features or intermediate representations from models such as XLS-R and WavLM for synthetic speech detection~\cite{yamaguchi2025investigating,tran2025multilevel}. Datasets such as CodecFake-Omni further emphasize the need to detect codec-based speech forgeries, where artifacts may be distributed across both local and contextual levels~\cite{du2025codecfake,chen2025source,du2025codecfakeplus}.

In speech generation, the connection to SSL mainly lies in learned acoustic units and pretrained speech representations. Recent TTS systems increasingly rely on discrete codec tokens, semantic speech units, or pretrained acoustic encoders as interfaces between speech signals and language-model-style generation. Models such as VALL-E 2 and WavTokenizer illustrate how learned audio units can support zero-shot speech synthesis and speaker adaptation~\cite{chen2024valle2,ji2025wavtokenizer}. Related codec and semantic-token approaches, including TS3-Codec and SemantiCodec, further show how compact audio representations can support efficient generative modeling~\cite{wu2024ts3,liu2024semanticodec}. Thus, in speech generation, SSL-related research contributes primarily through reusable acoustic units, semantic conditioning signals, and representation interfaces rather than through conventional feature extraction.

\subsection{Environmental Sound Analysis}
\label{subsec:env_apps}

Environmental sound analysis covers non-speech acoustic events from urban soundscapes and domestic environments to industrial machinery and natural scenes. Compared with speech, these signals are more heterogeneous, non-stationary, polyphonic, and sensitive to recording devices, acoustic spaces, and interference. The main challenge is not only label scarcity, but also the need to model open-ended event categories, overlapping sources, temporal uncertainty, and domain mismatch. SSL is well suited to this setting because masked modeling, contrastive learning, and token prediction can expose models to incomplete, corrupted, or weakly structured observations, encouraging representations that are less tied to fixed labels or specific recording conditions.

For general environmental audio representation learning, masked spectrogram modeling and discrete acoustic token prediction encourage models to infer missing or latent acoustic structures from surrounding time-frequency context. This matches the nature of environmental sounds, where event identity is often distributed across non-contiguous spectral and temporal regions. Models such as Audio-MAE~\cite{huang2022masked} and BEATs~\cite{chen2023beats} follow this direction, while Dasheng~\cite{dinkel2024scaling} further suggests that larger pretraining corpora and model capacity can improve transfer across heterogeneous audio tasks. Audio-text contrastive models such as CLAP~\cite{elizalde2023clap} address the open-vocabulary nature of environmental sounds by aligning audio with natural language descriptions, enabling zero-shot retrieval and recognition beyond fixed supervised label sets.

In sound event detection, SSL is particularly relevant because real acoustic scenes are often polyphonic and partially observed. Target events may overlap with other sources or be masked by background noise. Objectives based on masking, contrastive augmentation, or mix-based pretraining encourage the encoder to recover event-related structure from partial acoustic evidence. SSLAM~\cite{alex2025sslam}, for example, uses a mix-and-separate strategy to improve robustness to overlapping events. For spatially aware tasks such as sound event localization and detection, SSL features can be combined with multichannel spatial cues, as in STFF-Net~\cite{chen2024joint}, to jointly model event semantics and localization information.

For acoustic scene classification, the key issue is robustness to nuisance factors such as device response, background composition, and recording location. SSL can help by encouraging invariance to non-semantic perturbations through contrastive, augmentation-based, or teacher-student objectives. Recent work therefore uses pretrained SSL representations as transferable teachers or initialization models for compact scene classifiers, improving adaptation under limited labeled data and device constraints~\cite{cai2024leveraging}. Methods such as ECHO~\cite{Gupta2024echo} further introduce external semantic knowledge to guide scene representation learning, although such approaches are better viewed as weakly or semi-supervised extensions related to SSL rather than purely self-supervised methods.

Industrial anomalous sound detection provides another important use case. Since abnormal machine sounds are rare and often unavailable during training, the task naturally reduces to learning the acoustic structure of normal operation and detecting departures from it. Reconstruction-based, consistency-based, and domain-adaptive SSL objectives are well matched to this setting. Machine-aware adapters reduce source-target domain discrepancy~\cite{Han2025ExploringSA}, Anopatch~\cite{jiang2024anopatch} improves temporal embedding consistency under non-stationary noise, and DP-MAE~\cite{liu2024dp} uses masked reconstruction behavior to support anomaly scoring. These show that SSL benefits environmental sound analysis not simply by reducing annotation cost, but by matching the physical structure of acoustic scenes: overlap, sparsity, device variability, and deviation from normal patterns.

\subsection{Music Information Retrieval}
\label{subsec:mir}

Music information retrieval (MIR) differs from speech and environmental sound analysis in both signal structure and task definition. Musical signals are inherently polyphonic and hierarchical, with timbre, pitch, harmony, rhythm, style, and long-range form tightly coupled across multiple temporal scales. As a result, music representation learning must preserve fine-grained acoustic fidelity while also capturing higher-level musical semantics. This makes direct transfer from speech-oriented SSL models insufficient in many MIR tasks and has motivated music-specific pretraining objectives, tokenization strategies, and evaluation protocols.

A central issue in music SSL is the balance between acoustic detail and semantic abstraction. Speech-oriented units often emphasize intelligibility or phonetic content, whereas music representations must retain harmonic, melodic, and timbral information. Models such as MERT~\cite{li2024mert}, MuQ~\cite{zhu2025muq}, and MusicFM~\cite{won2024foundation} address this problem through music-aware tokenization and pretraining. MERT combines masked modeling over codec-derived tokens with CQT reconstruction to preserve melodic and harmonic structure~\cite{li2024mert}. MuQ introduces Mel-RVQ to improve the stability and expressiveness of music discretization~\cite{zhu2025muq}, while MusicFM explores large-scale music pretraining with BEST-RQ tokenization~\cite{won2024foundation}. Benchmarks such as MARBLE provide useful evaluation across music tagging, classification, retrieval, and structural analysis, helping assess whether learned representations capture both low-level and high-level musical properties~\cite{yuan2023marble}.

SSL is also valuable in MIR because many music labels are sparse, subjective, culturally dependent, or open-ended. Concepts such as mood, genre, instrumentation, performance style, and tonal system are difficult to exhaust with fixed closed-set labels. Audio-language contrastive models such as LAION-CLAP extend music understanding toward open-vocabulary retrieval and zero-shot annotation by aligning audio embeddings with natural language descriptions~\cite{laionclap2023}. In low-resource musical traditions, SSL representations can further support label propagation over representation manifolds. For example, graph-based transductive learning has been used to identify complex tonal structures such as ragas in Indian Art Music when labeled examples are limited~\cite{singh2025learning}. These cases show that SSL benefits MIR not only by reducing annotation cost, but also by supporting flexible semantic organization beyond fixed taxonomies.

For source separation and structural analysis, the advantage of SSL is tied to the layered and repetitive nature of music. Musical mixtures contain overlapping sources with correlated rhythms, harmonies, and timbres, while long-range forms such as verses, choruses, and transitions emerge over extended time scales. Methods such as MixIT~\cite{wisdom2020unsupervised} and Pac-HuBERT~\cite{chen2023pachubert} exploit mixture consistency or representation learning to separate sources without requiring fully isolated stems. For structural segmentation, changes in SSL embeddings across temporal scales can indicate transitions between musical sections, linking local acoustic variation to high-level form~\cite{buisson2022learning}.

Music generation is closely related to these representation trends, but it should not be conflated with SSL itself. Codec-language models such as MusicGen~\cite{copet2023simple} and controllable generation systems such as Music ControlNet~\cite{wu2024music} rely on learned audio tokens or structured intermediate representations to model long-range musical coherence and controllable attributes such as melody, rhythm, and dynamics. Their relevance to MIR lies in showing that tokenization and representation learning increasingly connect music understanding, retrieval, separation, and generation. Overall, SSL in MIR is most effective when it matches the multi-scale structure of music: local timbre and pitch, mid-level rhythmic and harmonic patterns, and long-range semantic or formal organization.

\subsection{Medical Audio and Bioacoustics: Learning Under Scarcity}
\label{subsec:med_bio}

Medical audio and bioacoustics are highly suitable application areas for SSL because they combine scarce expert annotations with rich acoustic regularities. In medical settings, labels often require clinical expertise, and recordings vary substantially across patients, devices, environments, and disease stages. In bioacoustics, annotation requires species knowledge and field observation, while target events may be rare or unevenly distributed across species. SSL is therefore useful not only for reducing annotation cost, but also for learning physiological or biological acoustic patterns that are difficult to specify through hand-crafted features.

In health-related acoustic analysis, the relevant information is often embedded in non-linguistic signal properties. Respiratory sounds, coughs, heart sounds, and sleep-related recordings contain periodicity, turbulence, abnormal events, and subject-specific variability. These structures align naturally with masked modeling, contrastive learning, and cross-modal SSL objectives, which encourage models to capture invariant or predictive acoustic patterns from unlabeled recordings. Health Acoustic Representations (HeAR), for example, are pretrained on large-scale respiratory and cough audio and transferred to downstream tasks such as tuberculosis screening and spirometry estimation~\cite{Baur2024HeARH}. This suggests that domain-specific SSL can capture physiological acoustic cues that are not adequately represented by general speech models. Related multimodal biosignal models, such as SleepFM~\cite{thapa2024sleepfm}, extend this idea by aligning EEG, ECG, and respiratory signals, while multi-view SSL methods learn representations that remain stable across heterogeneous sensor montages~\cite{yu2025multiview}. These studies are closely related to medical audio SSL, although they also belong to the broader area of self-supervised biosignal representation learning.

Speech-based health assessment provides another example where SSL matches the structure of the task. Neurological and psychiatric conditions may affect articulation, pause patterns, rhythm, speaking rate, voice quality, and prosody before these changes are fully reflected in lexical content. SSL speech encoders are useful here because they preserve multi-level acoustic and paralinguistic information from raw speech. wav2vec-style representations have been used to capture acoustic markers related to cognitive decline~\cite{luz2021detecting}, while DEPA uses a sequence-to-sequence pretraining objective to model long-term paralinguistic patterns for depression detection~\cite{zhang2021depa}. Recent LLM-assisted frameworks, such as ``Revise, Reason, and Recognize,'' combine acoustic cues, ASR outputs, and reasoning modules for speech-based mental health analysis~\cite{li2024revise}. These approaches should be interpreted as exploratory assessment tools rather than direct clinical decision systems, since clinical use requires validation across populations, recording conditions, and privacy-sensitive settings.

Hearing assistive technologies further illustrate the value of multimodal SSL. In degraded listening conditions or cochlear implant simulations, the acoustic signal may lose spectral detail that is important for intelligibility. Visual speech cues provide complementary articulatory information, making audiovisual SSL a physically meaningful strategy. Frameworks based on AV-HuBERT use visual speech information to guide the reconstruction or enhancement of degraded acoustic representations, showing that cross-modal pretraining can help compensate for missing or distorted auditory cues~\cite{lai2024leveraging}. The advantage here comes from matching the task requirement: when the acoustic channel is incomplete, another synchronized modality can provide constraints for representation learning.

In bioacoustics, SSL is valuable because many species have limited annotated data, but their vocalizations still contain recurring acoustic primitives such as pitch modulation, rhythm, call duration, spectral shape, and individual signatures. Masked prediction and representation transfer can exploit these regularities without requiring dense species-specific labels. AVES adapts a HuBERT-style masking objective to animal vocalizations and learns representations that generalize across taxa, including birds, bats, and primates~\cite{hagiwara2023aves}. Large-scale environmental audio pretraining can also support few-shot detection of rare species when only a small number of labeled examples are available~\cite{nolasco2023learning}. Evidence that speech-based SSL models such as WavLM encode individual vocal signatures in gibbon calls further suggests that some acoustic structures transfer across human and non-human vocal communication~\cite{cauzinille2024investigating}.

Overall, SSL benefits medical audio and bioacoustics when the target information is present in the signal but expensive to annotate explicitly. In medical audio, this includes physiological rhythms, pathological deviations, and paralinguistic markers; in bioacoustics, it includes species-level, event-level, and individual-level vocal patterns. The main open challenges are not only representation quality, but also domain-specific evaluation, robustness to recording variability, privacy protection, and careful validation before real-world clinical or ecological deployment.

\subsection{Multimodal Integration and Universal Audio Generation}
\label{subsec:multimodal_gen}

Multimodal audio learning extends SSL beyond single-modality acoustic representation learning by grounding audio in vision, language, and generative models. This direction is important because audio alone is often semantically ambiguous: the same acoustic event may correspond to different objects, actions, or intentions depending on visual or linguistic context. Multimodal SSL therefore helps not only by increasing pretraining data scale, but also by providing external semantic constraints that are unavailable from the audio waveform itself.

Audio-visual learning exploits the temporal correspondence between visible events and acoustic sources. This correspondence is physically meaningful because many sounds are produced by observable actions, object interactions, or scene dynamics. Early audio-visual SSL methods mainly used contrastive objectives to align global audio and visual embeddings. More recent models, such as CAV-MAE~\cite{gong2023contrastive} and MAViL~\cite{huang2023mavil}, combine contrastive learning with masked autoencoding, encouraging finer correspondence between audio and visual tokens. Such objectives are useful when the acoustic signal is incomplete or ambiguous, because visual context can help disambiguate sound sources and event semantics.

The scope of multimodal alignment has also expanded beyond paired audio-visual data. ImageBind~\cite{girdhar2023imagebind} projects audio and several other modalities into a shared embedding space anchored by vision, enabling zero-shot cross-modal transfer. This line of work shows that audio representations can be connected to broader multimodal spaces without relying solely on audio labels. Recent video-audio generation models such as JoVA~\cite{huang2025jova} further suggest that aligned multimodal representations can provide useful priors for temporally coherent generation. However, such systems are better viewed as SSL-adjacent multimodal foundation models rather than pure audio SSL methods.

Audio-language alignment addresses another limitation of closed-set audio classification. Many acoustic concepts are open-ended and are more naturally specified by language than by fixed labels. CLAP-style models align audio embeddings with text descriptions, enabling zero-shot retrieval, classification, and semantic search~\cite{elizalde2023clap}. In this setting, the advantage of SSL-related pretraining lies in mapping acoustic structure to a flexible semantic space, rather than simply learning a stronger audio encoder.

AudioLLMs further extend this idea by using pretrained audio encoders or SSL-derived acoustic representations as interfaces to large language models. Systems such as SALMONN~\cite{tang2024salmonn} and Qwen2-Audio~\cite{Qwen2-Audio} project speech, music, or environmental audio representations into language-model spaces, enabling instruction-following and reasoning over heterogeneous audio inputs. Omni-modal systems such as VITA~\cite{fu2025vita} and Mini-Omni~\cite{xie2024mini} further explore real-time or interactive multimodal processing. These systems should not be conflated with SSL itself; their relevance to this survey lies in showing how SSL-style acoustic pretraining and learned audio representations serve as front ends for broader audio-language intelligence.

Universal audio generation is another area where SSL-derived representations are increasingly used as semantic conditions, discrete units, or acoustic interfaces. AudioLDM 2~\cite{liu2024audioldm2}, for example, uses SSL-derived semantic representations to condition latent diffusion for text-to-audio generation. Token-based systems such as UniAudio~\cite{yang2024uniaudio} and Voicebox~\cite{le2024voicebox} reformulate multiple generation tasks as prediction over shared audio units, while flow-matching systems such as F5-TTS~\cite{chen2024f5} and F5R-TTS~\cite{sun2025f5r} improve efficiency and stability in speech generation. These models are not necessarily SSL methods in the narrow sense, but they are closely related to SSL through their use of learned acoustic units, pretrained encoders, or semantic audio representations.

Overall, multimodal integration and generation show how audio SSL is becoming connected to broader representation and foundation-model pipelines. Its main contribution in this setting is to provide acoustic representations that can be grounded in visual context, language descriptions, or generative objectives. The remaining challenges are substantial: multimodal models still struggle with temporal precision, controllability, hallucination, robustness under modality mismatch, and safety-sensitive audio content.

\begin{table*}[t]
\centering
\small
\caption{A comprehensive overview of benchmark suites and datasets for evaluating self-supervised audio representation learning.}
\renewcommand{\arraystretch}{1.18}
\resizebox{\linewidth}{!}{%
\begin{tabular}{p{4.2cm} p{9.0cm} p{4.0cm}}
\toprule
\textbf{Benchmark / Dataset} & \textbf{Evaluation Focus} & \textbf{Typical Metrics} \\
\midrule

\rowcolor[gray]{0.92}
\multicolumn{3}{l}{\textbf{Speech-oriented SSL Benchmark Suites (standardized multi-task evaluation)}} \\

SUPERB~\cite{yang2021superb} 
& Frozen-feature evaluation across ASR, phoneme/PR, keyword spotting, speaker ID/verification, emotion, etc. 
& WER/CER, Acc, EER, F1 \\

SUPERB-SG~\cite{tsai2022superb} 
& Extends SUPERB toward generative and semantic tasks (e.g., enhancement/separation/VC/ST variants depending on suite definition) 
& SI-SDR/SDR, PESQ/STOI, MCD, MOS, WER, task-specific \\

ZeroSpeech~\cite{nguyen2020zero} 
& Unsupervised/zero-resource speech discovery and modeling (challenge-style, rolling tasks) 
& ABX, clustering/token F1, bitrate, task-specific \\

LeBenchmark~\cite{evain2021task} 
& Standardized speech evaluation (e.g., French-centric, multilingual extensions depending on version) 
& WER, Acc, F1 \\

\rowcolor[gray]{0.92}
\multicolumn{3}{l}{\textbf{Low-resource Speech Pretraining Benchmarks}} \\

Libri-Light~\cite{kahn2020libri} 
& Pretraining with large unlabeled speech + limited labeled speech (typical 10min/1h/10h label settings) 
& WER/CER \\

\rowcolor[gray]{0.92}
\multicolumn{3}{l}{\textbf{Environmental and Non-speech Benchmarks}} \\

NOSS~\cite{shor2020towards} 
& Non-semantic speech tasks (speaker, emotion, language ID, etc.) often used for probing speech representations 
& Acc, macro-F1, EER \\

DCASE Tasks~\cite{mesaros2025decade} 
& ASC, SED, SELD, etc. with yearly standardized protocols 
& Acc, event-F1, ER, localization metrics \\

\rowcolor[gray]{0.92}
\multicolumn{3}{l}{\textbf{Universal Audio Representation Suites + Canonical General-Audio Benchmarks}} \\

HEAR~\cite{turian2022hear} 
& Unified evaluation of general audio embeddings across heterogeneous tasks/domains (frozen API-style evaluation) 
& Acc, mAP, task-specific \\

AudioSet~\cite{Gemmeke2017AudioSet} 
& Ontology-based multi-label audio event tagging; common for (zero-shot) audio-text alignment evaluation as well 
& mAP, AUC \\

FSD50K~\cite{fonseca2021fsd50k} 
& Open dataset for large-scale sound event tagging/classification 
& mAP, Acc \\

\rowcolor[gray]{0.92}
\multicolumn{3}{l}{\textbf{Audio-Text Multimodal SSL Benchmarks (audio-language alignment, retrieval, captioning)}} \\

AudioCaps~\cite{kim2019audiocaps} 
& Audio$\rightarrow$text caption generation; also used for evaluating audio-text alignment models via retrieval protocols 
& BLEU, METEOR, ROUGE-L, CIDEr, SPICE \\

Clotho~\cite{drossos2020clotho} 
& Captioning with emphasis on caption diversity; widely used benchmark for audio-language models 
& BLEU, METEOR, ROUGE-L, CIDEr, SPICE \\

WavCaps~\cite{mei2024wavcaps} 
& Large-scale weakly-labeled audio-text pairs for audio-language pretraining; evaluated on captioning/retrieval transfer 
& Retrieval R@K; Caption metrics (CIDEr, etc.) \\

\rowcolor[gray]{0.92}
\multicolumn{3}{l}{\textbf{Audio-Visual / Audio-Video Multimodal SSL Benchmarks (fusion, correspondence, grounding)}} \\

AVE~\cite{tian2018audio} 
& Temporal localization of audio-visual events (supervised/weakly-supervised, cross-modality localization) 
& mAP, localization Acc, IoU-based \\

VGGSound~\cite{chen2020vggsound} 
& Audio classification / AV correspondence from videos with visible sound source; common pretraining + evaluation benchmark 
& Top-1/Top-5 Acc \\

AVSBench~\cite{zhou2023avsbench} 
& Pixel-wise segmentation of sounding objects (single/multi-source settings) 
& mIoU, F-score \\

AVQA~\cite{yang2022avqa} 
& QA requiring joint audio-visual reasoning in dynamic scenes 
& QA Accuracy \\

\rowcolor[gray]{0.92}
\multicolumn{3}{l}{\textbf{Audio-Centered Multimodal Evaluation Suites (AudioLLM / Instruction / Robustness)}} \\

AudioBench~\cite{wang2025audiobench} 
& Universal benchmark for AudioLLMs (multi-task, multi-dataset): speech understanding, audio scene understanding, voice/paralinguistic understanding 
& Task-specific (Acc/F1/WER/CIDEr/mAP, etc.) \\

AV Robustness Bench (e.g., AVROBUSTBENCH)~\cite{maharana2025texttt} 
& Robustness evaluation under uni-/bi-modal corruptions and domain shifts for AV models 
& Robust Acc/mAP vs severity \\

\bottomrule
\end{tabular}
}

\label{tab:ssl_audio_benchmarks_big}
\end{table*}

\section{Benchmarking, Datasets, and Evaluation Protocols}
\label{sec:benchmarking}

Benchmarking defines how audio SSL representations are evaluated. As the field evolves, evaluation has expanded from speech-centric tasks to general-purpose and multimodal benchmarks covering diverse downstream scenarios.

\subsection{Benchmark Tasks for Audio SSL}
\label{subsec:benchmark_tasks}

\textit{a) Speech-centric Downstream Tasks:} Speech-related downstream tasks constitute the earliest and most extensively studied evaluation scenarios for audio SSL~\cite{huang2022investigating}. Automatic speech recognition, phoneme recognition, and keyword spotting primarily assess whether learned representations preserve fine-grained temporal and phonetic structures. Speaker identification and verification tasks, in contrast, emphasize the ability of representations to encode speaker-specific characteristics that are invariant to lexical content. Paralinguistic tasks, such as emotion recognition and language identification, further probe higher-level and longer-term acoustic attributes, providing complementary perspectives on representation quality.

\textit{b) Environmental Sound and Acoustic Scene Understanding:} Beyond speech, environmental sound and acoustic scene understanding tasks have become increasingly important for evaluating the generalization capability of audio SSL models~\cite{chandrakala2019environmental}. Acoustic scene classification focuses on global contextual cues over extended temporal windows, whereas sound event classification and detection require sensitivity to transient events and overlapping sound sources. Localization-oriented tasks, such as sound event localization and detection (SELD), additionally incorporate spatial information, thereby assessing whether representations can jointly capture temporal, spectral, and spatial characteristics of audio signals.

\textit{c) General Audio and Music-related Tasks:} General audio and music-related tasks further extend the evaluation scope toward broader semantic coverage~\cite{christodoulou2024multimodal}. Large-scale audio tagging benchmarks, including AudioSet~\cite{Gemmeke2017AudioSet} and FSD50K~\cite{fonseca2021fsd50k}, are commonly used to examine scalability and robustness under weakly labeled conditions. Music tagging and genre classification introduce domain-specific structures related to rhythm, harmony, and timbre. Cross-domain transfer tasks, which evaluate performance gaps between training and target domains, are particularly relevant for assessing the universality of learned audio representations.

\textit{d) Audio-centered Multimodal Tasks:} More recently, audio-centered multimodal tasks have emerged as an important evaluation frontier~\cite{guzhov2025audio}. Audio-text alignment tasks, such as retrieval and captioning, assess whether audio representations can be consistently mapped to semantic language spaces. Audio-visual understanding tasks, including event localization and segmentation, evaluate cross-modal correspondence and temporal synchronization. Audio-visual reasoning tasks, such as question answering, further challenge models to support higher-level semantic integration and reasoning across modalities.

\subsection{Representative Benchmark Datasets and Suites}
\label{subsec:benchmark_suites}

As summarized in Table~\ref{tab:ssl_audio_benchmarks_big}, the diversity of benchmark datasets reflects the evolving requirements of the field.

\textit{1) Speech Benchmark Suites and Challenges:} In the speech domain, standardized benchmark suites have substantially improved the comparability of audio SSL methods. SUPERB~\cite{yang2021superb} and its variants provide a unified multi-task evaluation framework that isolates representation quality through consistent downstream protocols. ZeroSpeech~\cite{bavin2015cambridge, dupoux2018cognitive, dunbar2022self} benchmarks focus on zero-resource settings, explicitly avoiding labeled supervision and emphasizing intrinsic representation properties. LeBenchmark~\cite{evain2021lebenchmark, parcollet2024lebenchmark} extends evaluation to multilingual scenarios, while Libri-Light~\cite{kahn2020libri} highlights low-resource pretraining conditions, reflecting realistic data availability constraints.

\textit{2) Non-speech and Environmental Audio Benchmarks:} For non-speech and environmental audio, the DCASE challenge series~\cite{mesaros2017dcase, mesaros2025decade} has become a de facto standard, covering tasks such as acoustic scene classification, sound event detection, and spatial audio analysis. Complementary to these semantic-oriented benchmarks, NOSS emphasizes non-semantic and paralinguistic speech attributes, thereby capturing aspects of audio representations that are less directly tied to lexical content.

\textit{3) Universal Audio Representation Benchmarks:} Universal audio representation benchmarks aim to evaluate generalization across heterogeneous domains. HEAR~\cite{turian2022hear} and HARES aggregate tasks from speech, environmental sound, and music into unified evaluation suites, providing a holistic view of representation transferability. Large-scale tagging datasets such as AudioSet~\cite{Gemmeke2017AudioSet} and FSD50K~\cite{fonseca2021fsd50k} remain central references for assessing performance under weak supervision and large vocabulary settings.

\textit{4) Audio-Text Multimodal Benchmarks:} In the audio-text multimodal domain, AudioCaps~\cite{kim2019audiocaps} and Clotho~\cite{drossos2020clotho} serve as standard benchmarks for audio captioning and retrieval. WavCaps~\cite{mei2024wavcaps} further supports large-scale audio-text pretraining and zero-shot evaluation. Retrieval-oriented protocols, typically measured using Recall@K and mean average precision, are commonly adopted to quantify semantic alignment between modalities.

\textit{5) Audio-Visual and Audio-Video Benchmarks:} Audio-visual and audio-video benchmarks further enrich the evaluation landscape. AVE focuses on audio-visual event localization, while VGGSound~\cite{chen2020vggsound} provides large-scale paired audio-visual data. AVSBench~\cite{zhou2023avsbench} targets sounding object segmentation, and AVQA~\cite{yang2022avqa} as well as MUSIC-AVQA evaluate multimodal reasoning capabilities. These benchmarks collectively emphasize temporal alignment, spatial correspondence, and cross-modal semantics.

\textit{6) Emerging AudioLLM Evaluation Suites:} With the emergence of Audio Large Language Models, new evaluation suites have been proposed to assess multi-task generalization and robustness. AudioBench~\cite{wang2025audiobench} offers a unified framework for evaluating diverse audio understanding tasks, while robustness-oriented benchmarks examine performance degradation under noise, distribution shifts, or adversarial perturbations.

\subsection{Evaluation Protocols}
\label{subsec:protocols}

\textit{1) Linear Probing and Frozen Representation Evaluation:} Evaluation protocols play a critical role in determining how representation quality is interpreted. Linear probing with frozen encoders remains a widely adopted protocol, particularly in benchmarks such as SUPERB and HEAR, as it minimizes the influence of downstream model capacity and optimization. This setting is often regarded as a proxy for the intrinsic separability of learned representations.

\textit{2) Full Fine-tuning Protocols:} Full fine-tuning protocols, in which the entire model is adapted end-to-end, are more common in large-scale tagging and multimodal tasks. While these protocols better reflect practical deployment scenarios, they introduce additional variability that can obscure direct comparisons between representation learning methods.

\textit{3) Low-resource and Semi-supervised Evaluation Settings:} Low-resource and semi-supervised evaluation settings restrict the amount of annotated downstream data, such as the 10-minute or 1-hour splits in Libri-Light, to assess the label efficiency of learned representations. These protocols align closely with a central motivation of SSL, namely reducing reliance on large-scale manual annotation. From an objective--architecture alignment perspective, low-resource performance is especially informative because it tests whether pretrained structural priors, such as local acoustic compression, sequential state propagation, or global contextual routing, can transfer to new tasks when downstream supervision is limited.

\textit{4) Zero-shot and Cross-modal Transfer Evaluation:} Low-resource and semi-supervised evaluation settings attempt to bridge these perspectives by restricting labeled data availability or adopting few-shot transfer scenarios. Such protocols align closely with the original motivation of SSL, namely reducing reliance on large-scale annotated datasets. Zero-shot and cross-modal transfer evaluations further extend this paradigm by measuring generalization without task-specific training, particularly in audio-text and audio-visual retrieval scenarios.

\begin{table*}[t]
\centering
\caption{Evaluation metrics for speech, audio, and multimodal generation tasks.}
\resizebox{1.0\linewidth}{!}{
\renewcommand{\arraystretch}{1.25}
\setlength{\tabcolsep}{6pt}

\begin{tabular}{p{0.14\linewidth} p{0.34\linewidth} c p{0.34\linewidth}}
\toprule
\textbf{Metric} & \textbf{Calculation / Methodology} & \textbf{Formula} & \textbf{Description} \\
\midrule

\rowcolor[gray]{0.92}
\multicolumn{4}{l}{\textbf{Speech Recognition and Generation Metrics}} \\

\textbf{WER} &
Align predicted and reference word sequences using dynamic programming. &
$ \mathrm{WER}=\frac{S+D+I}{N} $
&
Standard ASR metric measuring transcription accuracy. Lower is better. \\

\textbf{CER} &
Same as WER but computed at character level. &
$ \mathrm{CER}=\frac{S_c+D_c+I_c}{N_c} $
&
Useful for languages without explicit word boundaries. \\

\textbf{MOS} &
Human listeners rate samples; scores averaged across raters. &
$ \mathrm{MOS}=\frac{1}{M}\sum_{i=1}^{M} r_i $
&
Subjective evaluation of speech naturalness and quality. \\

\textbf{MCD} &
Mel-cepstral coefficients compared frame-wise and averaged. &
$ \mathrm{MCD}=\frac{10}{\ln 10}\sqrt{2\sum_{m}(c_m-\hat{c}_m)^2} $
&
Objective spectral distortion metric in speech synthesis. \\

\textbf{SI-SDR} &
Scale-invariant projection before distortion ratio calculation. &
$ \mathrm{SI\text{-}SDR}=10\log_{10}\frac{\lVert \alpha s\rVert ^2}{\lVert \alpha s-\hat{s}\rVert ^2} $
&
Robust metric for speech enhancement and separation. \\

\rowcolor[gray]{0.92}
\multicolumn{4}{l}{\textbf{Classification, Tagging, and Detection Metrics}} \\

\textbf{Accuracy} &
Compare predicted vs. ground-truth labels. &
$ \mathrm{Acc}=\frac{1}{N}\sum_{i=1}^{N}\mathbf{1}(y_i=\hat{y}_i) $
&
Overall correctness; sensitive to imbalance. \\

\textbf{Macro-F1} &
Compute per-class F1 then average. &
$ F1_{\mathrm{macro}}
=\frac{1}{C}\sum_{c=1}^{C}
\frac{2P_cR_c}{P_c+R_c} $
&
Balanced metric for imbalanced datasets. \\

\textbf{mAP} &
Average precision per class then mean over all classes. &
$ \mathrm{mAP}=\frac{1}{C}\sum_{c=1}^{C}\mathrm{AP}_c $
&
Standard metric for multi-label tagging tasks. \\

\textbf{Event-F1} &
Match detected vs. reference events with tolerance windows. &
- &
Captures temporal localization + classification accuracy. \\

\textbf{ER} &
Counts substitutions, deletions, insertions at event level. &
$ \mathrm{ER}=\frac{S+D+I}{N_{\mathrm{ref}}} $
&
Summarizes detection errors in a scalar form. \\

\rowcolor[gray]{0.92}
\multicolumn{4}{l}{\textbf{Verification and Representation Quality Metrics}} \\

\textbf{EER} &
Decision threshold where FAR equals FRR. &
$ \mathrm{FAR}=\mathrm{FRR} $
&
Threshold-independent speaker verification metric. \\

\textbf{ABX Error} &
Triplet discrimination probability in embedding space. &
$ P\big(d(A,X)>d(B,X)\big) $
&
Measures phonetic discriminability in zero-resource speech. \\

\rowcolor[gray]{0.92}
\multicolumn{4}{l}{\textbf{Multimodal Evaluation Metrics}} \\

\textbf{Recall@K} &
Check whether ground-truth appears in top-K ranked retrieval. &
$ R@K=\frac{1}{N}\sum_{i=1}^{N}
\mathbf{1}(\mathrm{gt}\in \mathrm{Top}\text{-}K) $
&
Used in audio-text and audio-visual retrieval tasks. \\

\textbf{BLEU} &
Modified n-gram precision with brevity penalty. &
- &
Surface-level similarity for caption generation. \\

\textbf{CIDEr} &
TF-IDF weighted n-gram consensus across references. &
- &
Emphasizes content relevance and agreement. \\

\textbf{SPICE} &
Scene-graph semantic proposition matching. &
- &
Evaluates semantic correctness beyond n-grams. \\

\textbf{QA Accuracy} &
Correct answers divided by total questions. &
$ \mathrm{Acc}=\frac{\#\mathrm{correct}}{\#\mathrm{questions}} $
&
Used in audio-visual reasoning benchmarks. \\

\textbf{mIoU} &
Mean IoU across all classes. &
$ \mathrm{mIoU}=\frac{1}{C}\sum_{c=1}^{C}
\frac{|P_c\cap G_c|}{|P_c\cup G_c|} $
&
Standard segmentation overlap metric. \\

\bottomrule
\end{tabular}}
\label{tab:metrics}
\end{table*}

\subsection{Evaluation Metrics}
\label{subsec:metrics}

\textit{a) Speech Recognition and Generation Metrics:} Evaluating speech-related models requires not only measuring transcription correctness but also assessing perceptual naturalness and signal-level fidelity. For intelligibility, \textbf{WER} and \textbf{Character Error Rate (CER)} remain the most widely adopted metrics, quantifying the normalized edit distance between predicted transcripts and ground-truth sequences. These metrics provide a direct reflection of how accurately generated or recognized speech preserves linguistic content, particularly in dialogue-oriented systems.

Beyond transcription accuracy, perceptual speech quality is commonly evaluated through subjective and objective criteria. \textbf{Mean Opinion Score (MOS)} serves as the gold-standard human evaluation protocol, where listeners rate the naturalness and overall quality of synthesized speech. Complementary to MOS, objective distortion metrics such as \textbf{Mel-Cepstral Distortion (MCD)} quantify spectral differences between reference and generated speech, and are widely used in speech synthesis benchmarks.

For speech enhancement and separation tasks, \textbf{Scale-Invariant Signal-to-Distortion Ratio (SI-SDR)} is adopted to measure reconstruction fidelity while eliminating scale ambiguity.

\textit{b) Classification, Tagging, and Detection Metrics:} In addition to generation quality, many multimodal systems require reliable classification and event-level detection capabilities. Standard metrics such as \textbf{Accuracy} measure the overall proportion of correctly predicted labels, while being sensitive to class imbalance.

To address imbalance and provide a more class-balanced evaluation, Macro-F1 computes the average F1 score across all classes, ensuring that minority categories contribute equally to the final assessment. For multi-label tagging scenarios, mean Average Precision (mAP) remains the dominant metric, evaluating ranking quality and confidence calibration across categories.

In temporal detection settings such as sound event detection, metrics extend beyond label correctness to localization accuracy. Event-F1 evaluates detection quality under onset/offset tolerances, while Error Rate (ER) summarizes insertion, deletion, and substitution errors at the event level, offering a unified scalar measure of detection robustness.

\textit{c) Verification and Representation Metrics:} For tasks involving speaker identity preservation and embedding robustness, verification-based metrics are essential. Equal Error Rate (EER) is widely used in speaker verification, representing the operating point where false acceptance and false rejection rates are equal. Its threshold-independent nature makes it particularly suitable for comparing representation quality across systems.

Furthermore, in zero-resource and phonetic discriminability studies, the ABX Error metric is employed to evaluate whether learned embeddings can correctly distinguish minimal acoustic contrasts. By comparing relative distances between triplet samples, ABX provides a fine-grained probe into the discriminative structure of speech representations.

\textit{d) Multimodal Metrics:} A defining characteristic of modern multimodal systems is the ability to align information across modalities such as audio, text, and vision. Retrieval-based metrics such as Recall@K are commonly used to assess cross-modal matching performance, measuring whether the correct counterpart appears among the top-ranked candidates.

For audio captioning and multimodal generation tasks, language-based metrics remain important. BLEU evaluates n-gram precision with brevity penalties, while CIDEr emphasizes consensus across multiple references through TF-IDF weighting. Beyond surface-level overlap, SPICE measures semantic correctness by comparing scene-graph propositions extracted from generated captions.

In multimodal reasoning benchmarks, QA Accuracy directly quantifies question-answer correctness, reflecting higher-level reasoning capabilities. Additionally, for spatial grounding and segmentation-related tasks, mean Intersection-over-Union (mIoU) is adopted to measure overlap consistency between predicted and ground-truth masks, serving as a standard metric for spatial understanding.

\section{Challenges and Limitations}\label{sec:challenges}

Despite the revolutionary progress in SSL and LALMs regarding feature extraction, cross-modal alignment, and generative tasks, the field still faces several deep-seated challenges in its pursuit of ``Universal Auditory Intelligence.'' A primary constraint is the \textbf{Tokenization Dilemma and Perceptual Bottleneck}. Current LALMs predominantly rely on neural audio codecs or SSL-based clustering to discretize continuous audio. This introduces a fundamental contradiction: capturing high-fidelity acoustic details requires high frame rates, which leads to a severe sequence length and frequency mismatch with the text modality. Conversely, highly compressed semantic tokens sacrifice fine-grained temporal structures, and these discrete tokenizers remain highly sensitive to imperceptible acoustic noise, where minor perturbations can cause drastic token flips—a phenomenon termed the ``perceptual bottleneck''.

Furthermore, there is a significant \textbf{Evaluation Bias and Global Pooling Bottleneck}. Traditional SSL evaluation paradigms, such as linear probing on frozen backbones, often fail to reflect true embedding quality because pre-training objectives like Masked Audio Modeling (MAM) tend to generate dispersed, context-dependent features. Conventional global pooling forcibly compresses these into a single vector, causing subtle yet critical sound events to be overshadowed by dominant signals. This necessitates a heavy reliance on computationally expensive full fine-tuning to achieve state-of-the-art performance.

The issues of Audio Hallucination and Text Prior Dependency also present major hurdles. Due to imbalanced data distributions and a heavy reliance on textual backbones, LALMs frequently suffer from ``perceptual hallucinations,'' where models prioritize learned textual priors over actual acoustic inputs. This leads to the fabrication of physically impossible events or the inversion of temporal causality. Coupled with this is the Security Vulnerability and Adversarial Attacks; audio models are exceptionally sensitive to adversarial noise. Attackers can bypass safety guardrails through signal-level manipulations like voice cloning, yet current safety alignment remains largely confined to the text level. Finally, a Catastrophic Forgetting and Spatio-Temporal Deficiency persists in continual learning scenarios. Most SSL frameworks collapse multi-channel audio into mono signals, discarding physical geometric features like Time Difference of Arrival (TDOA), which results in poor performance in deep spatio-temporal reasoning tasks such as 3D/4D sound localization.

\section{Future Trends}\label{sec:future_trends}

To address these limitations, the research trajectory for audio SSL and LALMs is evolving toward more unified and physics-aware architectures. A prominent trend is the adoption of \textbf{Continuous Representations and Flow Matching Paradigms}. To overcome the distortion and sparsity of discrete tokens, the field is migrating toward continuous or semi-discrete feature representations. Future architectures will likely utilize Flow Matching (e.g., F5-TTS) and Diffusion Transformers to resolve modality length mismatches while preserving high-fidelity paralinguistic information.

In parallel, the development of Audio Chain-of-Thought and Process Rewards is gaining momentum. Future models will evolve from ``thinking about audio'' to ``thinking with audio,'' gaining the ability to dynamically locate and re-listen to specific segments during reasoning. By integrating Reinforcement Learning (RL) with process-oriented rewards—inspired by System 2 reasoning models—researchers aim to incentivize logically consistent and hallucination-free multi-step reasoning.

Furthermore, the horizon is expanding toward Native Spatial Awareness and 4D Audio Intelligence. Future SSL models will move toward multi-channel architectures with native support for microphone arrays and binaural audio. By integrating physics-informed and geometry-grid representations, models will achieve universal sound source localization and motion trajectory reasoning, providing essential grounding for Embodied AI. Finally, Safety Alignment and Robust Defense Mechanisms will become central to pre-training. This involves disentangling identity from content in SSL representations and developing defensive strategies that encompass the acoustic layer of hearing to combat the proliferation of Deepfakes.

\section{Conclusion}\label{sec:conclusions}

This paper has examined audio self-supervised learning (SSL) from a structural alignment perspective that connects pretraining objectives, architectural inductive biases, and downstream applications. Rather than treating the field as a chronological progression of pretext tasks, we have emphasized how different supervisory signals create different processing demands, including local acoustic compression, invariance learning, sequential state propagation, contextual inference, semantic abstraction, and multimodal grounding. These demands help explain the shift from locality-biased encoders to Transformer-based global routing, SSM-based long-context modeling, and hybrid designs that combine complementary priors. Downstream tasks in speech processing, environmental sound analysis, music information retrieval, medical and bioacoustic analysis, and multimodal audio understanding provide practical tests of whether objective--architecture alignments generalize across domains. Challenges such as discrete tokenization, the perceptual bottleneck, long-context efficiency, and multimodal robustness suggest that further progress will require closer coordination among objectives, architectures, and application requirements.

\section{Acknowledgments}
This work is supported by National Science and Technology Major Project (2023ZD0121101).

% \section{Acknowledgments}
% This work is supported by National Science and Technology Major Project (2023ZD0121101), National University of Defense Technology (ZZCX-ZZGC-01-04) and Major Fundamental Research Project of Hunan Province (2025JC0005).

\bibliographystyle{IEEEtran}
\bibliography{references}

\end{document}